\documentclass[acmsmall, screen, 10pt, prologue,dvipsnames]{acmart}
\usepackage[vlined,ruled]{algorithm2e}
\usepackage{mathpartir}
\usepackage{algorithmic}
\usepackage[inline]{enumitem}
\usepackage{graphicx}
\usepackage{makecell}
\usepackage{colortbl}
\usepackage{mdframed}
\usepackage{mathtools}
\usepackage{wrapfig}
\usepackage{alltt}
\usepackage{lipsum}
\usepackage{wrapfig}
\usepackage{xspace}
\usepackage{subcaption}
\usepackage{multirow}
\usepackage{tikz-qtree}
\usepackage{xcolor}
\usepackage[compact]{titlesec}

\makeatletter
\newcommand{\ALG@noend}{\renewcommand{\alglinenumber}[1]{}\ALG@step}
\makeatother

\makeatletter
\let\@authorsaddresses\@empty
\makeatother

\setcopyright{rightsretained}
\acmDOI{10.1145/3632860}
\acmYear{2024}
\copyrightyear{2024}
\acmSubmissionID{popl24main-p83-p}
\acmJournal{PACMPL}
\acmVolume{8}
\acmNumber{POPL}
\acmArticle{18}
\acmMonth{1}
\received{2023-07-11}
\received[accepted]{2023-11-07}

\begin{document}

\author[N. Patton]{Noah Patton}
\affiliation{
  \position{Research Assistant}
  \department{Department of Computer Science}
  \institution{The University of Texas at Austin}
  \city{Austin}
  \state{Texas}
  \country{USA}
}
\email{npatt@cs.utexas.edu}
\orcid{0009-0002-7028-518X} 

\author[K. Rahmani]{Kia Rahmani}
\affiliation{
  \position{Postdoctoral Fellow}
  \department{Department of Computer Science}
  \institution{The University of Texas at Austin}
  \city{Austin}
  \state{Texas}
  \country{USA}
}
\email{kiar@utexas.edu}
\orcid{0000-0001-9064-0797}


\author[M. Missula]{Meghana Missula}
\affiliation{
  \position{Research Assistant}
  \department{Department of Computer Science}
  \institution{The University of Texas at Austin}
  \city{Austin}
  \state{Texas}
  \country{USA}
}
\email{meghanam@utexas.edu}
\orcid{0000-0002-1610-6198}

\author[J. Biswas]{Joydeep Biswas}
\affiliation{
  \position{Associate Professor}
  \department{Department of Computer Science}
  \institution{The University of Texas at Austin}
  \city{Austin}
  \state{Texas}
  \country{USA}
}
\email{joydeepb@utexas.edu}
\orcid{0000-0002-1211-1731}

\author[I. Dillig]{Işıl Dillig}
\affiliation{
  \position{Professor}
  \department{Department of Computer Science}
  \institution{The University of Texas at Austin}
  \city{Austin}
  \state{Texas}
  \country{USA}
}
\email{isil@cs.utexas.edu}
\orcid{0000-0001-8006-1230}

\newcommand{\todo}[1]{\textcolor{red}{{#1}}}
\newcommand{\new}[1]{{{#1}}}

\newcommand{\isil}[1]{\textcolor{orange}{{#1}}}

\newcommand{\edited}[1]{\textcolor{blue}{{#1}}}

\newcommand{\tool}[0]{\textsc{prolex}}
\newcommand{\bench}[0]{\textsc{prolex-ds}} 

\newcommand{\encode}[1]{{\mathtt{encode}({#1})}}

\newcommand{\mypar}[1]{\vspace{5pt}
\noindent
{\bf \emph{#1.}}
 }

\newcommand{\intt}[0]{\mathbb{Z}}   
\newcommand{\natt}[0]{\mathbb{N}}   
\newcommand{\booll}[0]{\mathbb{B}}   
\newcommand{\ALT}{\,\mid\,}

\newcommand{\seq}[1]{{#1}^*}
\newcommand{\set}[1]{\overline{#1}}
\newcommand{\stx}[1]{\color{stx_color}{\mathtt {#1}} \color{black}}
\newcommand{\demo}[1]{\color{black}{{#1}} \color{black}}

\newcommand{\cmt}[1]{\color{grey6}{\text{\texttt{}}} \color{black}}

\newcommand{\modf}[1]{\color{red}{{#1}} \color{black}}

\newcommand{\algcmt}[1]{\color{rule_label_color}{\# \text{{#1}}} \color{black}}

\newcommand{\prompt}[1]{\color{RedViolet}{{\textsf{#1}}} \color{black}}

\newcommand{\Dfrac}[2]{%
  \ooalign{%
    $\genfrac{}{}{1.4pt}0{#1}{#2}$\cr%
    $\color{white}\genfrac{}{}{0.6pt}0{\phantom{#1}}{\phantom{#2}}$}%
}

\newcommand{\locs}{\mathcal{L}}
\newcommand{\objs}{\mathcal{O}}
\newcommand{\cloc}{\ell}
\newcommand{\intp}{\mathcal{I}}
\newcommand{\demos}{\mathcal{D}}
\newcommand{\demosym}{\delta}


\newcommand{\env}[0]{\mathcal{E}}
\newcommand{\trace}[0]{t}
\newcommand{\absenv}{\hat{\env}}

\newcommand{\ie}{\textit{i.e.,}}
\newcommand{\eg}{\textit{e.g.,}}
\newcommand{\etc}{\textit{etc.}}

\newcommand{\jb}[1]{\textcolor{red}{Joydeep: {#1}}}
\newcommand{\secref}[1]{Section~\ref{sec:#1}}
\newcommand{\seclabel}[1]{\label{sec:#1}}

\newcommand{\RULE}[2]{\footnotesize
\frac{\begin{array}{c}#1\end{array}}
     {\begin{array}{c}#2\end{array}}}
\newcommand{\ruleLabel}[1]{
\begin{flushleft}
    {\scriptsize
    \textrm{\sc{\color{Brown} (#1)}}}
    \vspace{-1mm}
  \end{flushleft}}

\newcommand{\evalB}[2]{[\![#1]\!]_{#2}}

  \definecolor{rule_label_color}{rgb}{0, 0.3, 0.0}
  \definecolor{demo_color}{rgb}{0.6, 0, 0.0}
  \definecolor{stx_color}{rgb}{0, 0, 0.8}
\definecolor{cmt_color}{rgb}{0.5, 0.7, 0.3}

\definecolor{grey12}{rgb}{0.98,0.98,0.98}
\definecolor{grey11}{rgb}{0.96,0.96,0.96}
\definecolor{grey10}{rgb}{0.93,0.93,0.93}
\definecolor{grey9}{rgb}{0.9,0.9,0.9}
\definecolor{grey8}{rgb}{0.8,0.8,0.8}
\definecolor{grey7}{rgb}{0.7,0.7,0.7}
\definecolor{grey6}{rgb}{0.6,0.6,0.6}
\definecolor{grey5}{rgb}{0.5,0.5,0.5}
\definecolor{grey4}{rgb}{0.4,0.4,0.4}
\definecolor{grey3}{rgb}{0.3,0.3,0.3}
\definecolor{grey2}{rgb}{0.2,0.2,0.2}

\title{Programming-by-Demonstration for Long-Horizon Robot Tasks (Extended Version)}

\begin{abstract}
{\bf Abstract.}
The goal of \emph{programmatic Learning from Demonstration (LfD)} is to learn a policy in a programming language that can be used to control a robot's behavior from a set of user demonstrations. This paper presents a new
programmatic LfD algorithm that targets \emph{long-horizon robot tasks} which require synthesizing programs with complex control flow structures, including nested loops with multiple conditionals.
 Our proposed method  first learns a {program sketch}  that captures the target program's  control flow and then completes this sketch using an LLM-guided search procedure that incorporates a novel technique for proving unrealizability of programming-by-demonstration problems.
{We have implemented our approach in a  new tool called \tool\ and present the results of a comprehensive experimental evaluation on 120 benchmarks involving complex tasks and environments.
We show that, given a 120 second time limit, \tool\ can find a program
consistent with the demonstrations in 80\% of the cases. Furthermore,  for
{81\%} of the  tasks for which a solution is returned, \tool \ is able to find
the ground truth program with just one demonstration.
In comparison, CVC5, a syntax-guided synthesis tool, is only able to solve  25\% of the cases \emph{even when given the ground truth program sketch},
and an LLM-based approach, GPT-Synth, is unable to solve any of the tasks due to
the environment complexity. 
}
\end{abstract}

\addtolength{\abovecaptionskip}{-4pt}
\setlength{\textfloatsep}{8pt plus 1.0pt minus 2.0pt}
\setlength{\floatsep}{8pt plus 1.0pt minus 2.0pt}
\setlength{\dbltextfloatsep}{8pt plus 1.0pt minus 2.0pt}
\setlength{\dblfloatsep}{8pt plus 1.0pt minus 2.0pt}
\setlength{\abovedisplayskip}{1pt}



\maketitle
\section{Introduction}

Learning From Demonstration (LfD) is an attractive paradigm for teaching robots how to perform novel tasks in end-user environments~\cite{ARGALL2009469}.
While most classical approaches to LfD are based on black-box behavior cloning~\cite{9117169, gail}, recent work has argued for treating LfD as a program synthesis problem~\cite{ldips, https://doi.org/10.48550/arxiv.2303.01440,10.1145/3568162.3576991}.
In particular,  \emph{programmatic LfD} represents the space of robot policies in a domain-specific language (DSL) and learns a program that is consistent with the user's demonstrations.

Although this programmatic approach has been shown to offer several advantages over black-box behavior cloning in terms of data efficiency, generalizability, and interpretability~\cite{idips,
10.1145/3236386.3241340}, existing work in this space suffers from three key shortcomings: First, most prior techniques focus on simple  {Markovian} policies that select the \emph{next} action based only on the current state. As a result, the target programs have a simple decision-list structure, and the main difficulty lies in inferring suitable predicates for each branch.
Second, most existing techniques have only been applied to restricted domains with limited object and interaction types, such as robot soccer playing where the entities of interest are known a priori and comprise a small set.

Our goal in this work is to develop a programmatic LfD approach for \emph{long-horizon} tasks that commonly arise in service mobile robot settings --- \eg{} putting away groceries in a domestic setting, or proactively providing tools and parts to a mechanic in assistive manufacturing. Long-horizon robotics tasks are inherently more challenging, as the robot needs to reason about the interactions between a long \emph{sequence of actions} (\eg{} that the dishes must be cleared from a table before it can be wiped down) and the \emph{effect of specific environmental states} on sequences of actions (\eg{} a robot tasked with dusting a shelf must first remove all items from the shelf if it is not empty vs. directly dusting it if there are no items on it). This is in stark contrast to control tasks, such as motion control, where the policy just needs to select the next action for a single time step.

Recognizing the importance of long horizon tasks, recent work~\cite{10.1145/3568162.3576991} has proposed a multi-modal  user interface (combining natural language with hand-drawn navigation paths) to facilitate programmatic LfD in this setting. {This paper makes another stride towards that goal, but in an orthogonal direction, by learning more complex programs  from demonstration traces. In particular, the approach that we propose in this paper aims to (a) handle tasks that require complex control flow (such as nested loops with multiple conditionals) and (b) scale to demonstrations performed in complex environments with hundreds of objects and a large number of relationships to consider between those objects. }

\begin{figure}[t]
    \centering
\includegraphics[width=0.82\textwidth]{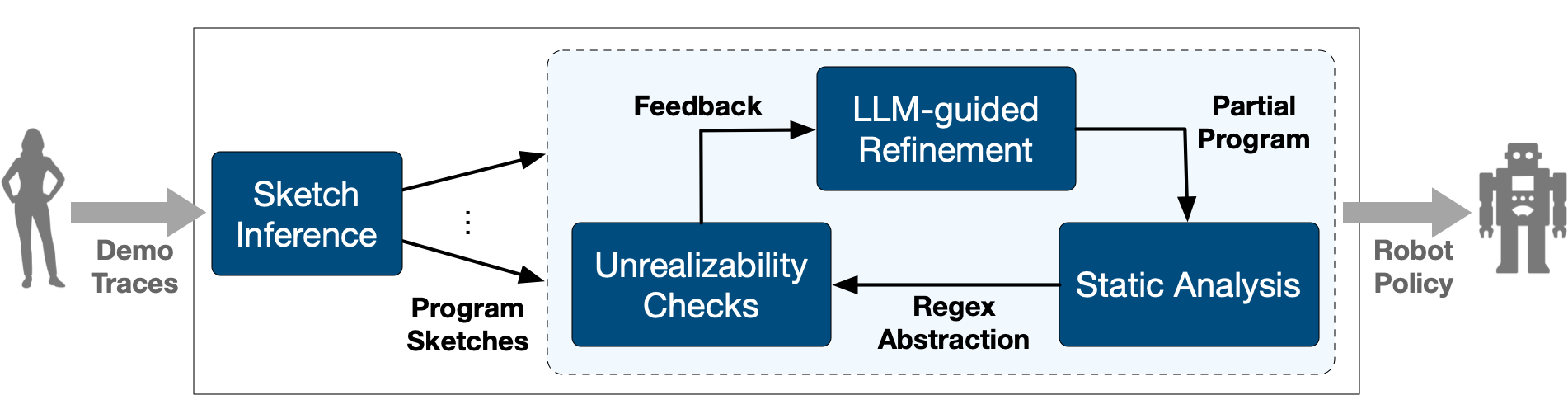}
    \caption{Overview of \tool}
    \label{fig:overview}
\end{figure}

Our proposed approach tackles  these two challenges using a novel program synthesis algorithm that is illustrated schematically in Figure~\ref{fig:overview}. First, given a set of demonstration traces, our approach infers a control flow sketch of the target program. To do so, our method abstracts each demo trace as a string over a finite alphabet and then learns a set of simple regular expressions that ``unify'' all of the demonstrations. As there is an obvious correspondence between regex operators and control flow structures (\eg{} loops as Kleene star; conditionals as disjunction), our method can quickly infer the control flow structure of the target program from a small number of demonstrations. Furthermore, because the sketch learner prefers small regular expressions over complex ones, this application of the Occam's razor principle introduces inductive bias towards  control structures that are more likely to generalize to unseen traces.

Given a program sketch capturing the underlying control flow structure, the second \emph{sketch completion} phase of our algorithm tries to find a complete program that is consistent with the given traces. This algorithm is based on top-down enumerative search, meaning that it starts by considering all  DSL programs as part of the search space and gradually refines it until it contains a single program. However, because the size of the search space is exponential with respect to the number of entities in the environment, such a search strategy does not scale to  complex environments. Our  approach deals with this challenge using two key ideas, namely (1) guiding the search  procedure using a large language model (LLM),  and (2) proving unrealizability of synthesis sub-problems.
\vspace{0.1in}

\noindent
{\bf LLM-guided refinement.}  As shown in Figure~\ref{fig:overview} (and as standard in the literature), our sketch completion procedure represents the search space  as a \emph{partial program} (\ie{} a program containing \emph{holes}), so the refinement step involves filling one of the holes in this partial program with a concrete expression. In our setting, these holes need to be instantiated with objects (or object types) in the environment, as well as properties of --and relationships between-- those objects. However, because the demonstration may be performed in complex environments with many such objects and properties, each hole typically has a very large number of possible completions. To address this problem,  our method consults an LLM to perform refinement; intuitively, this serves two purposes: First, when the target program contains object types that do not explicitly occur in the demonstrations (a very common scenario),  LLM guidance allows the synthesizer to propose new entities by reasoning about commonalities between objects  that do occur in the demonstrations. Second, by conditioning the current prediction on previous ones, the synthesizer can avoid generating programs that do not ``make sense" from a semantic perspective.
\vspace{0.1in}

\noindent
{\bf Proving unrealizability.} However, even with LLM guidance, the search procedure may end up constructing partial programs that have no valid completion with respect to the demonstrations. Our approach tries to avoid such dead-ends in the search space through a novel procedure for proving unrealizability. In particular, given a partial program $P$, our approach performs \emph{static analysis} to construct a suitable abstraction, in the form of a regex, that represents \emph{all} possible traces of \emph{all } completions of $P$ for the  demonstration environment. Given such a regex $r$, proving unrealizability of a synthesis problem boils down to proving that the demonstration trace cannot possibly belong to the language of $r$.

{We have implemented the proposed LfD technique in a tool called
\tool\footnote{\textbf{P}rogramming \textbf{RO}bots with \textbf{L}anguage
models and regular \textbf{EX}pressions} and evaluate it on a benchmark set
containing 120 long-horizon robotics tasks involving household activities. Given
a 2 minute time limit, our approach can complete {80\%} of the synthesis tasks
and can handle tasks that require multiple loops with several conditionals as
well as environments with up to \emph{thousands} of objects and  dozens of
object types. Furthermore, for {81\%} learning tasks that \tool\ is able to
complete within the 2 minute time limit, \tool\ learns a program that matches
the ground truth  from just a \emph{single} demonstration.
To put these results in context, we compare our approach against two relevant
baselines, including CVC5, a state-of-the-art SyGuS solver and GPT-Synth, a
neural program synthesizer, and experimentally demonstrate the advantages of our
approach over other alternatives. CVC5 is only able to solve 25\% of
the tasks even when given the ground-truth sketch. 
{GPT-Synth, on the other hand, is unable to solve any of the tasks due to environment complexity, even when the environment is simplified to include only  a small fraction of the objects in addition to the required ground truth entities.
}
%
%
%
Furthermore, we report the results of a series of ablation
studies and show that our proposed ideas contribute to successful synthesis.  }


In summary, this paper makes the following contributions:
\begin{itemize}[leftmargin=*]
    \item We propose a novel programming-by-demonstration (PBD) technique targeting  long-horizon robot tasks. Our approach can learn programs with complex control flow structures, including nested loops with conditionals, from a small number of traces and in complex environments with thousands of objects.
    \item We propose a new (reusable) method for proving  unrealizability of synthesis problems in the PBD setting.


    \item We implement these ideas in a tool called \tool\ and  evaluate its efficacy in the context of 120 benchmarks involving 40 unique household chores in three  environments.  \tool\ can complete the synthesis task within 2 minutes for {80\%} of the benchmarks, and, for 81\% of  the completed tasks,  \tool \ is able to learn the ground truth program from just a \emph{single} demonstration. \\

\end{itemize}



 \vspace{-0.1in}
 \vspace{-0.1in}

\section{Motivating Example}
\label{sec:mot}

\begin{figure}[t]
\begin{subfigure}[b]{0.42\textwidth}
\centering
\includegraphics[width=\textwidth]{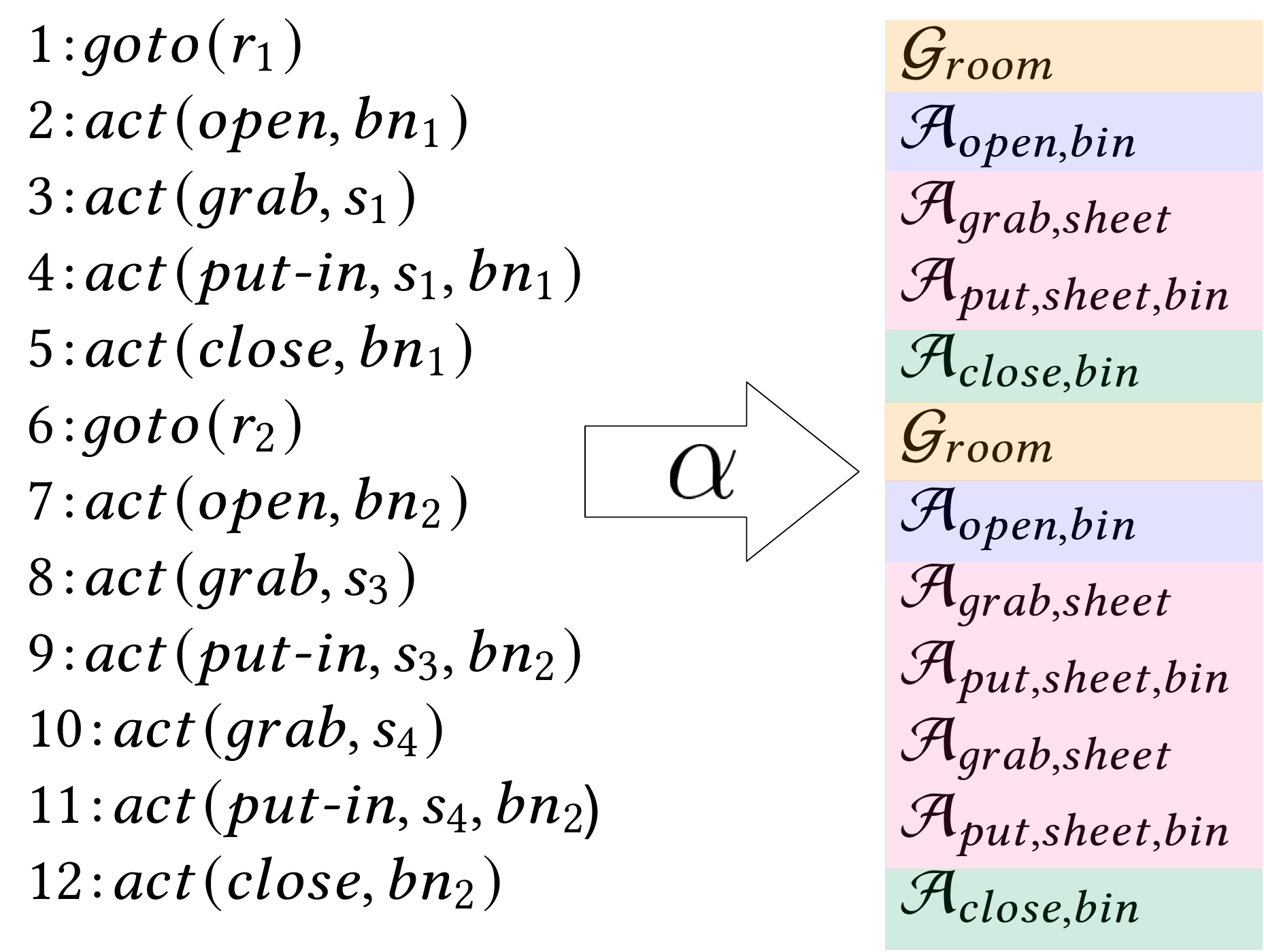}
\caption{Trace of demonstrated actions (left) and the abstracted string using function $\alpha$ (right)}
\label{subfig:demo-and-abs}
\end{subfigure}
\hfill
\begin{subfigure}[b]{0.52\textwidth}
\centering
\includegraphics[width=\textwidth]{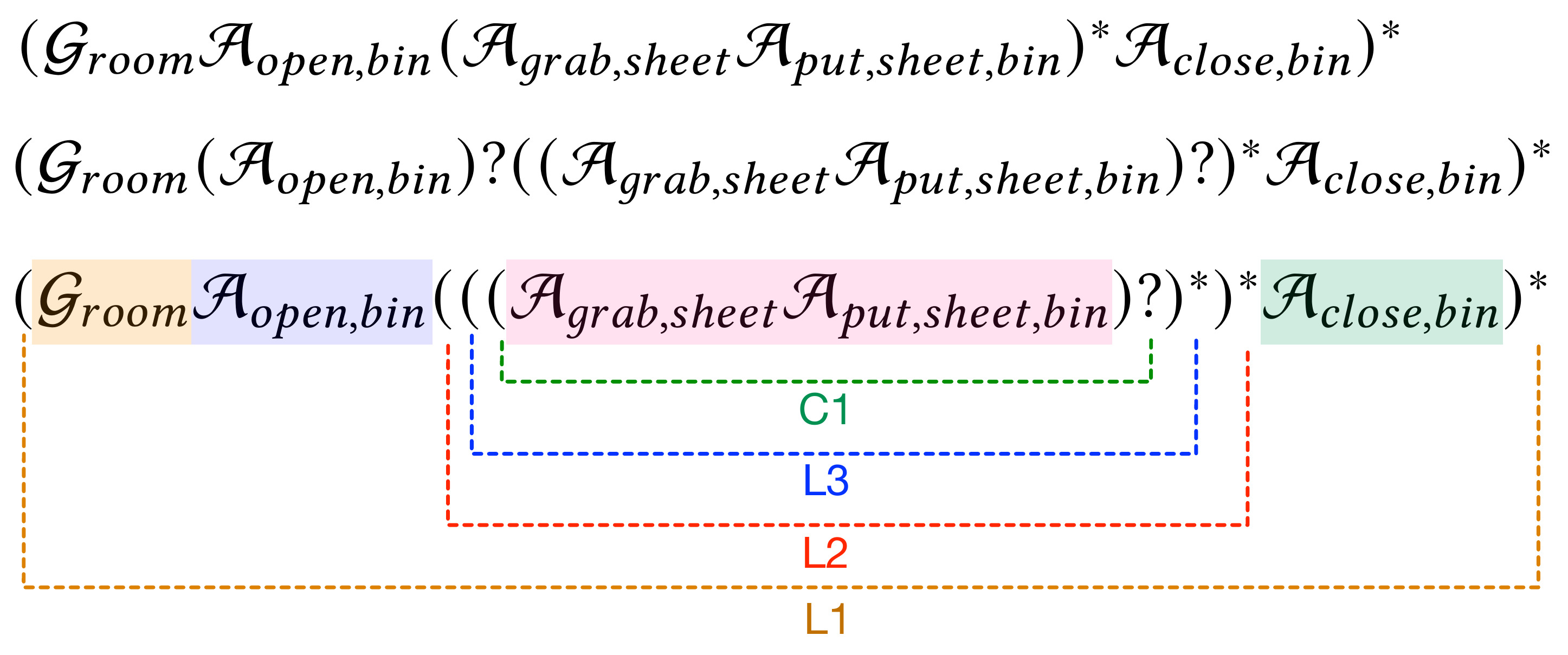}
\caption{Regexes learned from the  string abstracted from the demonstration. The correct regex (bottom) is highlighted to show looping and conditional structures.}
\label{subfig:regexes}
\end{subfigure}
\\[5mm]
\begin{subfigure}[b]{0.47\textwidth}
\centering
\includegraphics[height=38mm]{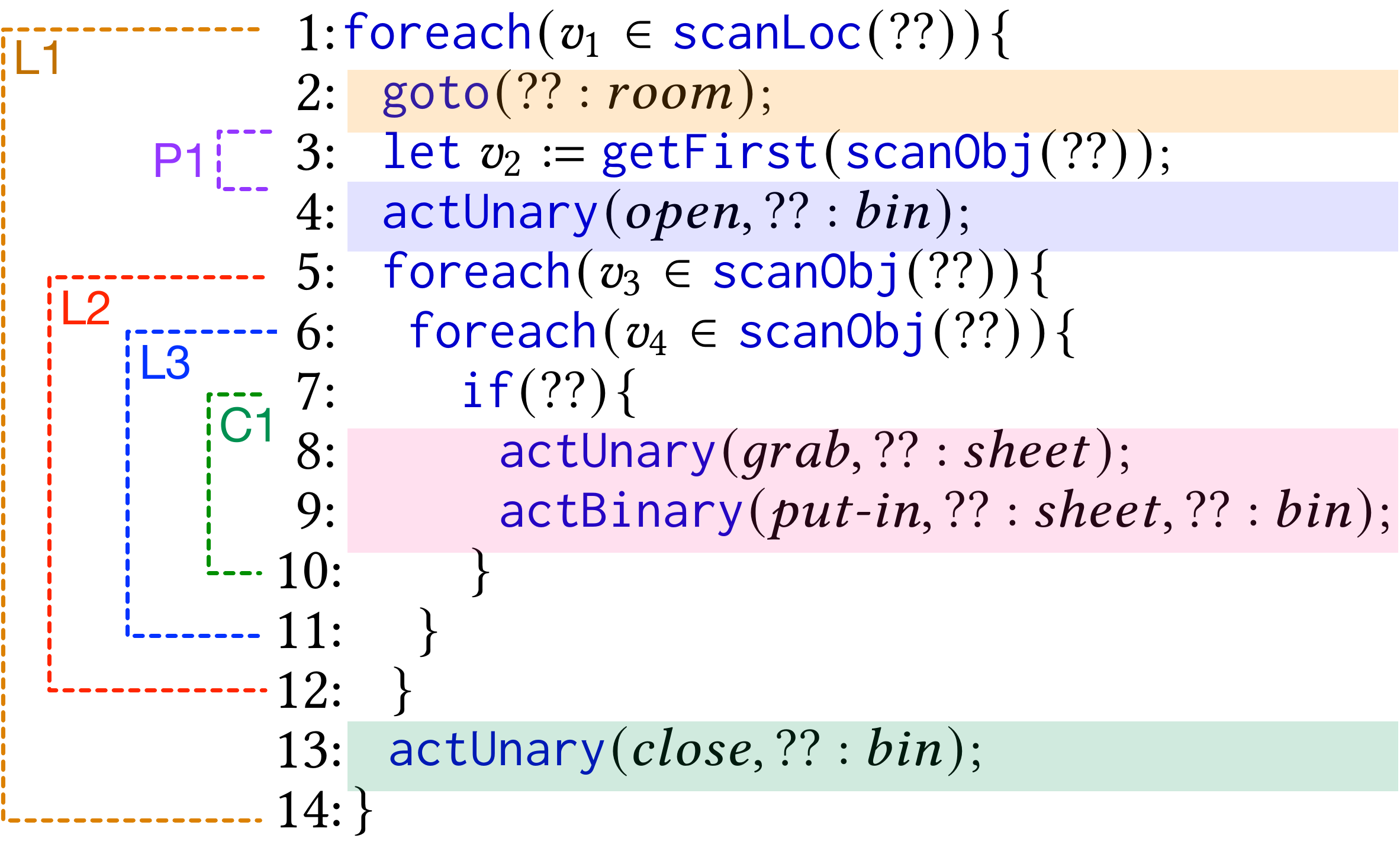}
\caption{Inferred Sketch}
\label{subfig:sketch}
\end{subfigure}
    \hfill
    \begin{subfigure}[b]{0.51\textwidth}
\centering
\includegraphics[height=38mm]{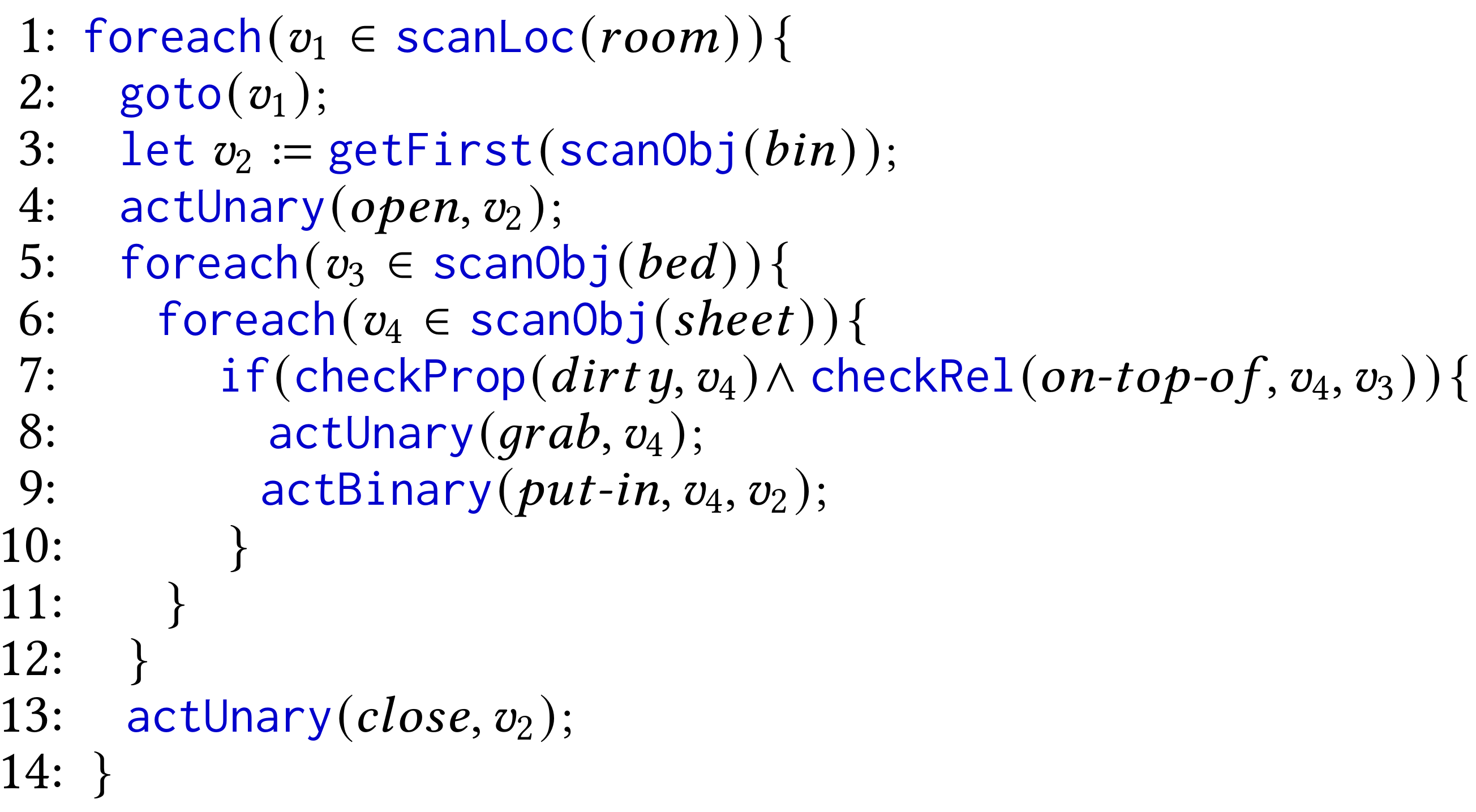}
\caption{Synthesized Program}
\label{subfig:synthesizer_program}
\end{subfigure}
    \caption{Motivating Example}
    \label{fig:motiv-ex}
\end{figure}


In this section, we present a motivating example to illustrate  our programmatic LfD approach.
%
Imagine a hotel worker who wants to instruct a robot  to collect dirty sheets from guest rooms and place them in a laundry bin in the room. 
The goal of LfD is to teach this task  through demonstrations rather than explicitly programming the robot.
We formalize user demonstrations in a form that is amenable to be captured using smart hand-held devices, similar to existing end-user robots like the iRobot Roomba~\cite{irobot}  and Amazon Astro~\cite{Lee2023}.


\begin{wrapfigure}[9]{r}{0.5\textwidth}
    \centering
    \scriptsize
\begin{tabular}{|c|l|l|}
\hline 
\textbf{Loc} &\multicolumn{2}{c|}{\textbf{State}}   \\ \hline \hline  
\multirow{3}{*}{$r_1$}    &
\(
    {Objs}\) &\( bed\!:\![b_1],\; sheet\!:\![s_1, s_2],\; bin\!:\![bn_1],\dots  \)\\ \cline{2-3} &
  \( { Props }\)&\(
   (s_1,dirty),(bn_1,closed)
   ,\; \dots\)  \\ \cline{2-3} &
   \( {Rels}\)&\(
(s_1,on\text{-}top\text{-}of,b_1),\; (s_2,on\text{-}top\text{-}of,b_1)
    ,\;\dots 
\)  
\\ \hline  \hline  
\multirow{3}{*}{$r_2$}    & \(
    {Objs}\)&\( bed\!:\![b_2],\; sheet\!:\![s_3, s_4],\; bin\!:\![bn_2],\dots \) \\ \cline{2-3} &
    \({Props}\)&\( 
    (s_3,dirty),(s_4,dirty),(bn_2,closed)
    ,\;\dots \) \\ \cline{2-3} &
   \( {Rels}\)& \(
    (s_3,on\text{-}top\text{-}of,b_2),\; (s_4,on\text{-}top\text{-}of,b_2)
    ,\;\dots  
\)     \\ \hline  
\end{tabular}
\caption{Partial representation of the initial environment. 
}
\label{subfig:demo_envs}
\end{wrapfigure}
\mypar{User Demonstration}
For the above task, suppose that the hotel worker performs a demonstration
consisting of the 12 actions shown in the left side of Figure~\ref{subfig:demo-and-abs}.
The demonstration takes place in two rooms, $r_1$ and $r_2$; Figure~\ref{subfig:demo_envs} shows the state of these two rooms before the demonstration takes place. Each room contains a large number of objects, including a bed, a laundry bin, and a few sheets on the bed.
In particular, there is a clean sheet ($s_2$) and a dirty sheet ($s_1$) on the bed in $r_1$, and there are two dirty sheets ($s_3$ and $s_4$) on the bed in $r_2$.
Note that the complete representation of the rooms includes  many more object types and properties, which are omitted from the figure due to space constraints.
The first five steps of the demonstration sequence shown in Figure~\ref{subfig:demo-and-abs} (left) correspond to the actions performed in $r_1$, and the remaining seven steps indicate the actions performed in $r_2$.
Specifically,
$\demo{goto(l)}$  indicates going to the location $l$, and $\demo{act(a, \overline{o})}$ indicates performing a specific action $a$ on objects $\overline{o}$.
Hence, in our example demonstration, the user  first visits room $r_1$, where they open bin $\demo{bn_1}$, grab and place sheet $\demo{s_1}$ in that bin, and finally close the bin. Next, they go to the second room, $\demo{r_2}$, and repeat a similar sequence of actions with the bin $\demo{bn_2}$, and sheets $\demo{s_3}$ and $\demo{s_4}$.








\mypar{Desired Output} Since our goal is to perform \emph{programmatic} LfD, we wish to learn a programmatic \emph{robot execution policy}  from the provided demonstration.
Figure~\ref{subfig:synthesizer_program} shows the  desired policy that generalizes from the user's only demonstration.  Intuitively, this policy encodes that the robot should go to each room (line~2), identify a bin and all beds present in that room, collect all \emph{dirty} sheets from the top of each bed (lines~7-8), and place them in the bin (line~9).  The program also contains \emph{perception} primitives that enable the robot to become aware of its environment (lines~1, 3, 5, and 6).
In particular, the function $\stx{scanObj}(\tau)$ allows the robot to identify all objects of type $\tau$ that are visible from its current location and reason about their properties and relations.
Function $\stx{scanLoc}(\tau)$ is similar but returns all locations of type $\tau$.
Synthesis of appropriate perception primitives is crucial for effective LfD, since the robot must be able to observe and reason about the state of objects in new and unseen environments.



\mypar{Synthesis Challenges} In this context, generating the desired policy from the user's demonstration is challenging for several reasons:
(1) the desired program has complex control flow, with three nested loops and an if statement with multiple predicates, (2)  the desired program requires performing appropriate perception actions that have no correspondence in the given demonstrations, and (3) the desired program requires reasoning about high-level concepts that are also not indicated in the demonstrations, such as being dirty or being on top of some other object. {Additionally, observe that the synthesized program needs to refer to object types (e.g., bed) that are not involved in the demonstration. Hence, the synthesizer cannot \emph{only} consider those objects in the demonstration, as the desired program could refer to any of the entities in the environment. This means that the difficulty of the synthesis task is inherently sensitive to the complexity of the environment. }

\mypar{Our Approach}
Our algorithm first generates a set of \emph{sketches} of the target program based on the demonstration. These sketches capture the  control flow structure of the target program but contain many missing expressions (``holes"). The goal of the subsequent sketch completion step is to fill these holes in a way that scales to complex environments.

\mypar{Sketch generation} To generate a sketch, our method first \textit{abstracts} the user's demonstrations as a set of strings:   Figure~\ref{subfig:demo-and-abs} (right) shows the  string abstracted from the user's sole demonstration using an abstraction function $\alpha$, where $\mathcal{G}$ and $\mathcal{A}$ denote $\demo{goto}$ and $\demo{act}$ in the demonstration, and the subscripts indicate the \emph{type} of their arguments. For example, even though the demonstration specifies that the user grabbed specific sheets (namely, $\demo{s_1}$, $\demo{s_3}$, and $\demo{s_4}$), the string abstraction omits such details and represents all three instances of this action using $\mathcal{A}_{grab,sheet}$. Each character in the abstract string is highlighted using a different color to visually aid the reader. The idea of converting the  demonstration to a more abstract form is a  crucial first step towards generalization.

Next, our approach utilizes existing techniques to synthesize a regex that matches the string encoding of the demonstrations:
Figure~\ref{subfig:regexes} presents three regular expressions that all match the string abstraction of the given demonstration.
Note that there is an obvious parallel between regex operators and the program's control flow: Since Kleene star denotes repetition, it naturally corresponds to a looping construct. Similarly, since the optional operator (i.e., $(r)?$) denotes choice, it is naturally translated into an $\stx{if}$ statement.
Thus, given a candidate regex $r$ for the demonstrations, our approach translates it into a sketch in a syntax-directed way. For instance, the third regular expression from Figure~\ref{subfig:regexes} is translated to the sketch shown in Figure~\ref{subfig:sketch}, where
the three nested loops and the conditional block are
marked using the same colored dashed lines both in the regex and in the sketch.
Additionally, because the program should not manipulate any objects before it {perceives} them, our sketch generation procedure also inserts any necessary perception primitives to the sketch. 
For instance, the   sketch in Figure~\ref{subfig:sketch} contains an inferred let binding (labelled \color{Fuchsia}{$\mathsf{P1}$}\color{black}), since the object of type $bin$ at line 4 cannot be acted upon before it is first perceived by the robot.



 \mypar{Sketch completion} For each sketch generated in the first step, our method tries to find a completion that matches the user's demonstration. In practice, several of these sketches are unrealizable, meaning that there is no completion that will match the user's demonstrations. For example, consider the first  regex presented above in Figure~\ref{subfig:regexes}
 that  matches the string abstraction of the demonstration. This regex does not include the optional operator $?$ used in the correct regex. Intuitively, sketches generated from this regex will be infeasible because, without an if statement in the sketch, the resulting programs would end up grabbing \emph{both} sheets in room $r_1$, rather than only the single dirty sheet. Hence, the second phase of our technique considers multiple sketches and tries to find a completion of \emph{any} sketch that is consistent with the given demonstrations.

Our sketch completion algorithm is based on top-down enumerative search but (a) utilizes a novel unrealizability checking procedure to quickly detect dead-ends and (b) leverages an LLM to guide exploration. In particular, starting from a sketch, the synthesis algorithm maintains a worklist of partial programs containing holes to be filled. When dequeuing a partial program from the worklist, we consider its probability according to the LLM, so more promising partial programs are prioritized over less likely ones. Going back to our example, consider a partial program where we test the color of a bedsheet before putting it in the bin. Because the concepts ``laundry bin'' and ``cleanliness'' are much more related compared to  ``color'' and ``laundry bin'', our technique prioritizes a program that includes the conditional $\stx{checkProp}(dirty, ...)$ over one that is based on color.

\begin{wrapfigure}{r}{0.46\textwidth}
    \centering
    \includegraphics[height=42mm]{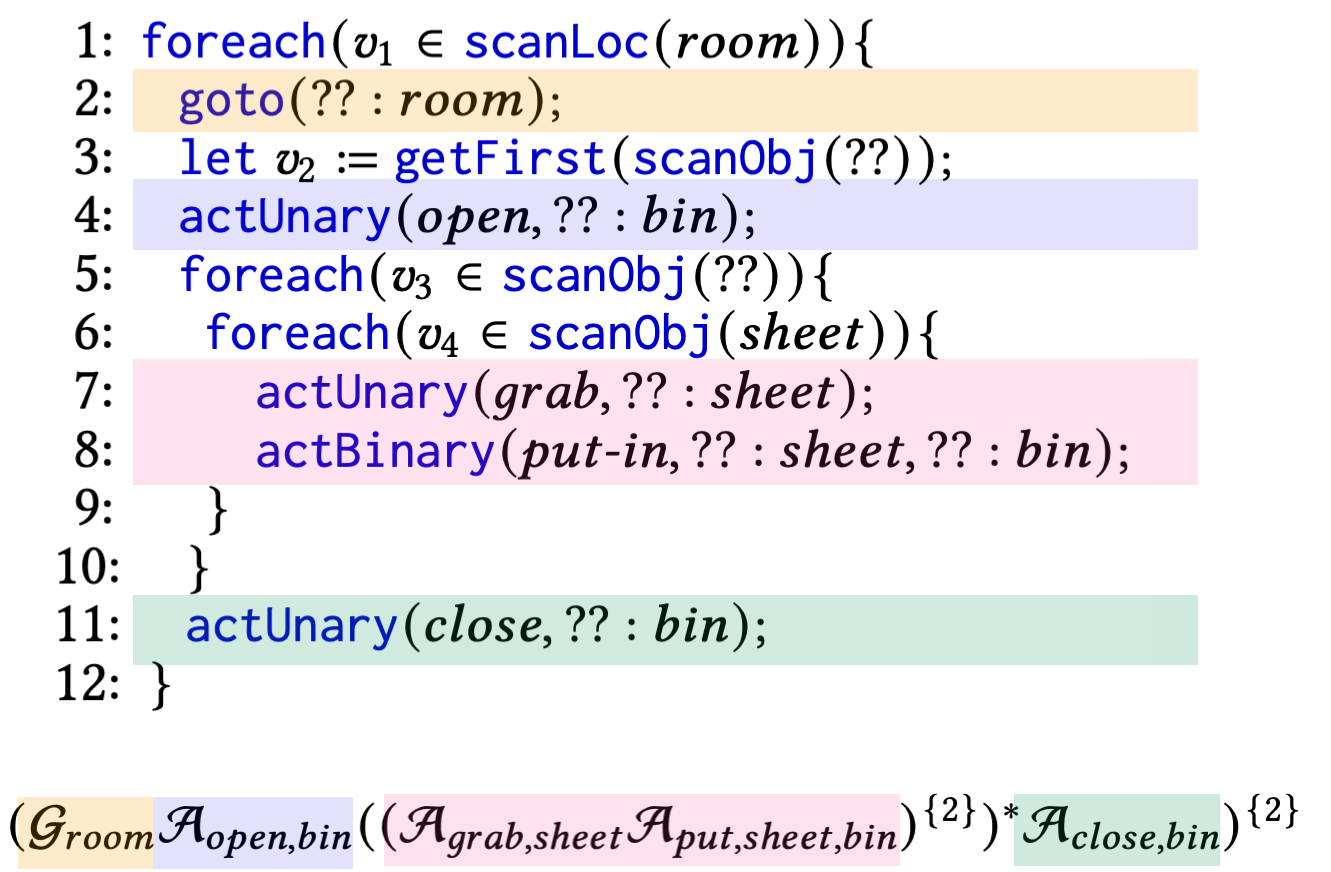}
    \caption{An Unrealizable Partial Program}
    \label{fig:unreal_regex}
\end{wrapfigure}

Our sketch completion method also leverages the demonstration to prove unrealizability of a given synthesis problem. As an example, consider  the partial program shown in Figure~\ref{fig:unreal_regex}. We can prove that there is no completion of this partial program that will be consistent with the user's demonstrations: Because none of the actions in the partial program modify object locations and because there are two sheets in each room, any completion of this partial program would end up performing the $\demo{grab}$ action \emph{at least four times} but there are only three $\demo{grab}$ actions in the trace. Hence, this partial program is a dead end, and our approach can detect unrealizability of such a synthesis sub-problem. To do so, it statically analyzes the partial program to infer upper and lower bounds on the number of loop executions. It then uses this information to construct a regex, shown at the bottom of Figure~\ref{fig:unreal_regex}, that summarizes all possible traces that this partial program can generate. Since the demo trace from Figure~\ref{fig:motiv-ex}(a) is not accepted by the regular expression in Figure~\ref{fig:unreal_regex}, our algorithm can prove the unrealizability of this synthesis problem.

\def\scalefactor{0.3 }
\begin{figure}[t]
\footnotesize
    \begin{mathpar}
  \begin{array}{lllll}
\tau_l & \in & \text{{Location\,Type}}^* & := & \{{bedroom},\; {kitchen},\; {basement},\; {mail room},\;\dots\} 
\\
\tau_o & \in & \text{{Object\,Type}}^* & := & \{ {plate},\; {drawer},\; {dishwasher},\; {sink},\;\dots\}  
\\
p & \in & \text{{Property}}^* & := & \{ {empty},\; {broken},\; {green},\; {dry},\;\dots\}  
\\
r & \in & \text{{Relation}}^* & := & \{ {inside\text{-}of},\; {next\text{-}to},\; {on\text{-}top\text{-}of},\; {belongs\text{-}to},\;\dots\} 
\\
 a & \in &  \text{{Action}}^* & := &\{{grab},\;  {open},\;  {pour\text{-}into},\;  {scrub\text{-}with},\;  {put\text{-}in},\; \dots \} 
\\
\rho & \in & \text{Item List} & := &  \stx{scanObj}(\tau_o) \ALT \stx{scanLoc}(\tau_l) 
    \\
   \phi & \in & \text{Conditional} & := & \stx{checkProp}(p,v) \ALT \stx{checkRel}(r,v_1, v_2) \ALT \phi_1 \wedge \phi_2 \ALT \phi_1 \vee \phi_2 \ALT \neg \phi
    \\
    \pi & \in & \text{Program} & := &  \stx{actUnary}(a,v) \ALT \stx{actBinary}(a,v_1,v_2)  \ALT \stx{goto}(v)\ALT \stx{if}(\phi) \{\pi\} \ALT \stx{skip} \ALT  
\\
& & & &
\stx{foreach}(v \in \rho) \{\pi\} 
  \ALT
\stx{let} \;v:=\stx{getNth}(\rho,n)\ALT
\pi;\pi
    
\end{array}
  \end{mathpar}
  \caption{DSL syntax where $v$ denotes variables and $n$ is a natural number. Rules marked with $^*$ domain-specific. }
  \label{fig:syntax}
\end{figure}

\section{Robot Execution Policies}
\label{sec:progs}

In this section, we introduce a domain-specific language (DSL) for programming robots and provide a formal definition of the robot learning from demonstrations (LfD) problem. 

\vspace{-0.04in}
\subsection{Syntax}
\label{subsec:syntax}
\vspace{-0.04in}

The syntax of the our DSL is presented in Figure~\ref{fig:syntax}.
A robot program ($\pi$) contains functions to perform various operations on a single object ($\stx{actUnary}$) or a pair of objects ($\stx{actBinary}$). The robot can move between locations using the $\stx{goto}$ function. The robot becomes aware of its  location by scanning the environment using $\stx{scanLoc}$, and  it becomes aware of objects in its current location using the $\stx{scanObj}$ function. The result of running a scan operation is an ordered list of location or object instances (denoted by $\rho$) of the specified type $\tau$. For example, $\stx{scanObj}(plate)$ yields all plates at the current location of the robots. Specific elements in the scan result can be bound to variables using a restricted  {\tt let} binding of the form $ {\tt let}  \ v := \stx{getNth}(\rho, n)$. This expression  introduces a new variable $v$ and assigns the $n$'th element of list $\rho$ to $v$. As standard, the DSL also contains typical conditional and looping control structures. Conditional expressions  check properties of objects ($\stx{checkProp}$), relationships between them ($\stx{checkRel}$), as well as their Boolean compositions. 
{
The abstractions for robot actions and perception in our DSL is similar to widely used accepted symbolic abstractions for classical planning~\cite{aeronautiques1998pddl,fox2003pddl2}, and more recently, symbolic robot policies~\cite{code-as-policies}.
}

Note that the DSL presented  in Figure~\ref{fig:syntax} is parametrized over a set of domain-specific terminals, indicated by an asterisk.  For example, location types $\tau_l$  are not fixed and can vary based on the target application domain. For example, for robot execution policies targeting household chores, locations might be kitchen, living room, basement etc. Similarly, object types, properties, relations, and actions are also domain-specific and can be customized for a given family of tasks.



\subsection{Operational Semantics}
\label{subsec:opsem}

In this section, we present the operational semantics of our robot DSL using the small-step reduction relation $\Rightarrow$ shown in Figure~\ref{fig:operational_semantics}. This relation  formalizes how the robot interacts with its environment while executing the program.
Specifically, the relation $\Rightarrow$ is defined between tuples of the form $(\pi,\env,\sigma,\trace)$, where $\pi$ is a program, $\env$ is the robot's execution environment, $\sigma$ is a valuation (mapping variables to their values), and  $\trace$ is a program trace.  In more detail, the environment $\env$ is a quadruple $(\locs, \objs, \cloc, \intp)$ where $\locs$ is a set of typed location identifiers; 
$\objs$ is a mapping from (typed) object identifiers to their corresponding location; $\cloc$ is the current location of the robot; and $\intp$ is an interpretation for all the relation symbols. That is, for a  relation $p$, $\intp(p)$ yields the set of tuples of objects for which $p$ evaluates to true.
Given object type $\tau_o$ and a location $l$, we write $\env.{\tt objs}(l,\tau_o)$ to denote the list of all objects that are at location $l$ and that have type $\tau_o$. Similarly, given a location type $\tau_l$, the list of locations of this type is denoted by $\env.\mathtt{locs}(\tau_l)$.
 Finally, a trace is a sequence of actions performed by the robot. Robot actions are denoted using $\demo{act(a,\bar{o})}$ and $\demo{goto(l)}$, where $a$ is an action that was performed on objects $\bar{o}$, and $l$ is a location that the robot visited. Note that robot execution policies are effectful programs:  for example, the location of the robot or some properties of an object can change after executing $\pi$. 

  \def\horizontalSpace{2mm}
\begin{figure}[t]
\footnotesize
\begin{minipage}[t]{0.3\textwidth}
  \ruleLabel{sequence}
$$
\RULE{
\pi,\env,\sigma,\trace
\Rightarrow 
\pi',
\env',
\sigma',
\trace'
}
{
\pi;\pi'',\env,\sigma,
\trace \Rightarrow 
\pi';\pi'',
\env',
\sigma',
\trace'
}
$$
\end{minipage}
\qquad \qquad
\begin{minipage}[t]{0.27\textwidth}
  \ruleLabel{skip}
$$
\RULE{
\\
}
{
\stx{skip};\pi,
\env,
\sigma,
\trace
\Rightarrow 
\pi,
\env,
\sigma,
\trace
}
$$
\end{minipage}
\\[\horizontalSpace]
\begin{minipage}[t]{0.3\textwidth}
  \ruleLabel{if-t}
$$
\RULE{
\phi \Downarrow_{\env,\sigma}\top 
}
{
\stx{if}(\phi)\{\pi\},
\env,\sigma,
\trace
\Rightarrow 
\pi,
\env,\sigma,
\trace
}
$$
\end{minipage}
\qquad \qquad 
\begin{minipage}[t]{0.27\textwidth}
  \ruleLabel{if-f}
$$
\RULE{
\phi\Downarrow_{\env,\sigma}\bot
}
{
\stx{if}(\phi)\{\pi\},
\env,\sigma,
\trace \Rightarrow 
\stx{skip},
\env,\sigma,
\trace
}
$$
\end{minipage}
\\[\horizontalSpace]
\begin{minipage}[b]{0.28\textwidth}
  \ruleLabel{act-unary}
$$
\RULE{
\pi=\stx{actUnary}(a,v)
\\
{o} = \sigma(v)\qquad
\env \xrightarrow{a, {o}} \env'
\\
\trace' = \demo{act(a,{o})}
}
{
\pi,
\env,\sigma,
\trace
\Rightarrow 
\stx{skip},
\env',\sigma,
t;t'
}
$$
\end{minipage}
\hfill
\begin{minipage}[b]{0.34\textwidth}
  \ruleLabel{act-binary}
$$
\RULE{
\pi=\stx{actBinary}(a,v_1,v_2)
\\
{o}_1 =\sigma(v_1)
\qquad
{o}_2 =\sigma(v_2)
\\
\env\xrightarrow{a,{o}_1,{o}_2} \env'
\qquad 
\trace'=\demo{act(a,{o}_1,{o}_2)}
}
{
\pi,
\env,\sigma,
\trace
\Rightarrow 
\stx{skip},
\env',\sigma,
t;\trace'
}
$$
\end{minipage}
\hfill
\begin{minipage}[b]{0.29\textwidth}
  \ruleLabel{goto}
$$
\RULE{
\pi=\stx{goto}(v)
\\
{l} = \sigma(v)\qquad
\env'=\env[\ell\mapsto {l}]
\\
\trace'=\demo{goto({l})}
}
{
\pi,
\env,\sigma,
\trace
\Rightarrow 
\stx{skip},
\env',\sigma,
t;\trace'
}
$$
\end{minipage}
\\[\horizontalSpace]
\begin{minipage}[b]{0.34\textwidth}
  \ruleLabel{let-obj}
$$
\RULE{
\pi = \stx{let} \;v:=\stx{getNth}(\stx{scanObj}(\tau_o),n)
\\
\sigma' = \sigma[v\mapsto \env.\mathtt{objs}(\ell,\tau_o) [n]]
}
{
\pi,
\env,
\sigma,
\trace
\Rightarrow \stx{skip},\env,\sigma',\trace
}
$$
\end{minipage}
\hfill
\begin{minipage}[b]{0.57\textwidth}
  \ruleLabel{foreach-obj}
$$
\RULE{
 n = |\env.\mathtt{objs}(\ell,\tau_o)|
 \qquad 
\forall_{0\leq i < n}.\; \sigma_i = \sigma[v\mapsto \env.\mathtt{objs}(\ell,\tau_o)[i]]
\\
\env_0 = \env 
\qquad
\trace_0 = \trace
\qquad
\forall_{0\leq i < n}.\;
\pi,
\env_i,
\sigma_i, 
\trace_i
\Rightarrow 
\stx{skip},
\env_{i+1},
\_\,,
\trace_{i+1}
}
{
\stx{foreach}(v \in \stx{scanObj}(\tau_o)) \{\pi\},
\env,
\sigma, 
\trace
\Rightarrow 
\stx{skip},
\env_{n},
\sigma, 
\trace_{n}
}
$$
\end{minipage}
\\[\horizontalSpace]
\begin{minipage}[b]{0.35\textwidth}
  \ruleLabel{let-loc}
$$
\RULE{
\pi =\stx{let} \;v:=\stx{getNth}(\stx{scanLoc}(\tau_l),n)
\\
\sigma' = \sigma[v\mapsto \env.\mathtt{locs}(\tau_l)[n]]
}
{
\pi,\env,\sigma,\trace \Rightarrow \stx{skip},\env,\sigma',\trace
}
$$
\end{minipage}
\hfill
\begin{minipage}[b]{0.57\textwidth}
  \ruleLabel{foreach-loc}
$$
\RULE{ 
n = |\env.\mathtt{locs}(\tau_l)|
\qquad
\forall_{0\leq i < n}.\; \sigma_i = \sigma[v\mapsto \env.\mathtt{locs}(\tau_l)[i]]
\\
\env_0 = \env 
\qquad
\trace_0 = \trace
\qquad
\forall_{0\leq i < n}.\;
\pi,
\env_i,
\sigma_i, 
\trace_i
\Rightarrow 
\stx{skip},
\env_{i+1},
\_\,,
\trace_{i+1}
}
{
\stx{foreach}(v \in \stx{scanLoc}(\tau_l)) \{\pi\},
\env,
\sigma, 
\trace
\Rightarrow 
\stx{skip},
\env_{n},
\sigma, 
\trace_{n}
}
$$
\end{minipage}
\caption{Operational semantics. Relation $\Downarrow$ is defined in Figure~\ref{fig:bool_expr_reduction}, and the definition of relation $\rightarrow$ can be found in Appendix ~\ref{app:aux_rel}.}
\label{fig:operational_semantics}
\end{figure}


  With the above notations in place, we now explain the operational semantics from Figure~\ref{fig:operational_semantics} in more detail. 
The first rule, labeled \textsc{(sequence)}, defines how the robot takes a step by executing the first statement in the program. The next rule (\textsc{skip}) defines the semantics of executing a $\stx{skip}$ statement, which has no effect on the execution state. 
The rules \textsc{(if-t)} and \textsc{(if-f)} describe the flow of the program when a conditional statement $\stx{if}(\phi)\{\pi\}$ is executed. First, the Boolean expression $\phi$ is evaluated to $\bot$ or $\top$, depending on the result, the program either skips or executes $\pi$. Boolean expressions are evaluated using the relation $\Downarrow$ defined in Figure~\ref{fig:bool_expr_reduction}. This relation is parameterized by the execution environment $\env$ and  valuation $\sigma$. 

\def\horizontalSpace{2mm}
\begin{figure}[t]
\footnotesize
\begin{minipage}[t]{0.23\textwidth}
  \ruleLabel{check\_prop\_t}
$$
\RULE{
\textbf{o} \in \intp(p) 
}
{
\stx{checkProp}(p,v) \Downarrow_{\env,\sigma} \top 
}
$$
\end{minipage}
\hfill
\begin{minipage}[t]{0.23\textwidth}
  \ruleLabel{check\_prop\_f}
$$
\RULE{
\sigma(v) \not\in \intp(p) 
}
{
\stx{checkProp}(p,v) \Downarrow_{\env,\sigma} \bot 
}
$$
\end{minipage}
\hfill
\begin{minipage}[t]{0.25\textwidth}
  \ruleLabel{check\_rel\_t}
$$
\RULE{
(\sigma(v_1),\sigma(v_2)) \in \intp(r)
}
{
 \stx{checkRel}(r,v_1,v_2) \Downarrow_{\env,\sigma} \top
}
$$
\end{minipage}
\hfill
\begin{minipage}[t]{0.25\textwidth}
  \ruleLabel{check\_rel\_f}
$$
\RULE{
(\sigma(v_1),\sigma(v_2)) \not\in \intp(r)
}
{
 \stx{checkRel}(r,v_1,v_2) \Downarrow_{\env,\sigma} \bot
}
$$
\end{minipage}
\\[\horizontalSpace]
\begin{minipage}[t]{0.14\textwidth}
  \ruleLabel{negation}
$$
\RULE{
\phi\Downarrow_{\env,\sigma} b
}
{
\neg \phi \Downarrow_{\env,\sigma} \neg b
}
$$
\end{minipage}
\qquad
\begin{minipage}[t]{0.26\textwidth}
  \ruleLabel{conjunction}
$$
\RULE{
\phi_1\Downarrow_{\env,\sigma} b_1
\qquad
\phi_2\Downarrow_{\env,\sigma} b_2
}
{
\phi_1\wedge \phi_2\Downarrow_{\env,\sigma} b_1\wedge b_2
}
$$
\end{minipage}
\qquad
\begin{minipage}[t]{0.26\textwidth}
  \ruleLabel{disjunction}
$$
\RULE{
\phi_1\Downarrow_{\env,\sigma} b_1
\qquad
\phi_2\Downarrow_{\env,\sigma} b_2
}
{
\phi_1\vee \phi_2\Downarrow_{\env,\sigma} b_1\vee b_2
}
$$
\end{minipage}
\caption{Boolean expression evaluation.
}
  \vspace{-0.05in}
\label{fig:bool_expr_reduction}
\end{figure}

The \textsc{(act-unary)} and \textsc{(act-binary)} rules 
 specify the outcomes of executing a unary and a binary action, respectively, using the auxiliary relation $\rightarrow\,\subseteq\env\times\env$. Given an environment $\env$, an action $a$ and affected object instance(s), the relation $\rightarrow$  formalizes how the environment $\env$ is modified based on the semantics of  action $a$. Since our DSL is parameterized over the set of actions,  we do not discuss the $\rightarrow$ relation in detail in the main body of the paper and refer the interested reader to  Appendix~\ref{app:aux_rel} for a representative subset of actions used in our evaluation.

The \textsc{goto}
rule   defines the effect of executing a $\stx{goto}(v)$ statement, where the environment is updated to reflect the robot's new location, and a new trace element $\demo{goto({l})}$ is generated and appended to the existing trace, where $l$ is the location stored in variable $v$.
Next, the rules \textsc{let-obj} and \textsc{let-loc}, define the  semantics of the $\stx{let}\;v:=\stx{getNth}(\rho,n)$ statement, which assigns to variable $v$ the $n^{th}$ element of list $\rho$ obtained via either $\stx{scanObj}$ or $\stx{scanLoc}$. Specifically,  $\stx{scanObj}(\tau_o)$ yields all objects of type $\tau_o$ that are present at the robot's current location, and  $\stx{scanLoc}(\tau_l)$ yields all locations of type $\tau_l$.  
The rules (\textsc{foreach-obj}) and (\textsc{foreach-loc}) describe the semantics of loops of the form $\stx{foreach}(v\in \rho)\{\pi\}$, where $\rho$ is the result of a scan operation. As expected, these rules iteratively bind $v$ to each of the elements in $\rho$ and execute the loop body $\pi$ under this new valuation. 



 Finally, we use the $\Rightarrow$ relation to define the semantics of executing a policy $\pi$ on environment $\env$. Given  $\env$ and robot execution policy $\pi$, we write $\pi(\env) = \trace$ iff $
 \pi, \env, \mathsf{Nil}, \mathsf{Nil}  \Rightarrow \stx{skip}, \_, \_, \trace$
 where $\mathsf{Nil}$ denotes an empty list/mapping.

\subsection{Problem Statement}

In this section, we formally define the  LfD problem that we address in this paper. Informally, given a set of demonstrations $\demos$, the LfD problem is to find a robot execution policy $\pi^*$ (in the DSL of Figure~\ref{fig:syntax}) such that $\pi$ is consistent with $\demos$. To make the notion of consistency more precise, we represent a demonstration $\demosym$ as a pair $(\env, \trace)$ where $\env$ is the initial environment and $\trace$ is a trace of the user's demonstration in this environment. 

\begin{definition} \label{def:completed}{\bf (Consistency with demonstration)} We say that a robot execution policy $\pi^*$ is consistent with a demonstration $\demosym = (\env, \trace)$, denoted $\pi \models \demosym$, iff $\pi(\env) = \trace$.  
\end{definition}

We also extend this notion of consistency to a set of demonstrations $\mathcal{D}$, and we write $\pi \models \mathcal{D}$ iff $\pi \models \delta$ for every demonstration $\delta$ in $\mathcal{D}$. We can now formalize our problem statement as follows:

 \begin{definition}
 \label{def:solved}
 {\bf (Programmatic LfD)} Given a set of demonstrations $\demos$, the \emph{programmatic LfD} problem is to find a robot execution policy $\pi^*$ such that $\pi^* \models \mathcal{D}$.
\end{definition}

\section{Synthesis Algorithm}
\label{sec:synth}
In this section, we present our synthesis technique for solving the programmatic LfD problem defined in the previous section. We start by giving an overview of the top-level algorithm and then describe each of its key components in more detail.

\SetKw{KwInput}{Input:}
\SetKw{KwProc}{Procedure:}
\SetKw{KwOutput}{Output:}
\SetKw{KwReturn}{return}
\SetKw{KwCont}{continue}

{
\begin{algorithm}[t]
\scriptsize
\caption{Top-level Synthesis Algorithm }\label{alg:top}
\KwInput{A set of demonstrations $\mathcal{D}$, a statistical completion model $\theta$
}\\
\KwOutput{A policy consistent with the demonstrations or $\bot$ if none exists}\\
\begin{algorithmic}[1]
\STATE $\mathsf{Synthesize}(\mathcal{D}, \theta)$
\STATE \ \ \ \ $A := \{\alpha(t) \ALT (\_\,,t)\in \mathcal{D}\}$ \ \ \ \algcmt{get the abstraction of given traces using function $\alpha$}
\vspace{2pt}
\STATE \ \ \ \ {\bf while} ($true$) \{
\STATE \ \ \ \ \ \ \ \ $r := \mathsf{GetNextRegex}(A)$ \ \ \ \algcmt{get a regular expression that matches the abstractions of all demos}
\STATE \ \ \ \ \ \ \ \ {\bf if}$(r=\bot) \  \ \mathbf{break}$
\STATE \ \ \ \ \ \ \ \ {\bf while} ($true$) \{
\STATE \ \ \ \ \ \ \ \ \ \ \ \ $s := \mathsf{GetNextSketch}(r)$  \ \ \ \algcmt{get the next lazily generated sketch from $r$}
\STATE \ \ \ \ \ \ \ \ \ \ \ \ {\bf if}$(s=\bot) \ \  \mathbf{break} $
\STATE \ \ \ \ \ \ \ \ \ \ \ \ $\pi:= \mathsf{CompleteSketch}_\theta(s,\mathcal{D})$ \ \ \ \algcmt{search for a consistent completion of $s$ and return the result}
\STATE \ \ \ \ \ \ \ \ \ \ \ \ {\bf if}$(\pi\not\eq\bot)$ {\bf return} $\pi$ 
\STATE \ \ \ \ \ \ \ \ \}
\STATE \ \ \ \ $\}$
\STATE \ \ \ \  {\bf return }$\bot$
\end{algorithmic}
\end{algorithm}
}

\subsection{Top-Level Algorithm}

Our top-level learning procedure is presented in Algorithm~\ref{alg:top}. This algorithm takes as input a set of demonstrations $\mathcal{D}$ and returns a policy $\pi$ such that for all $\delta \in \mathcal{D}$, we have $\pi \models \delta$. If there is no programmatic policy that is consistent with all demonstrations, the algorithm returns $\bot$.

The synthesis procedure starts by  constructing an \emph{abstraction} of  each demonstration $\delta \in \mathcal{D}$ as a string over the  alphabet $\Sigma = \{ \mathcal{G}_\tau, \mathcal{A}_{a, \tau}, \mathcal{A}_{a, \tau, \tau'} \} 
$
where $\tau, \tau'$ indicate location and object types (e.g., ${plate}$, ${kitchen}$) and $a$ denotes a specific type of action (e.g., ${grab}$). This abstraction is performed at line 2 of Algorithm~\ref{alg:top} using the function $\alpha$, defined as follows: 
\[
\begin{array}{lll}
\alpha(\demo{goto(l)})  :={\mathcal{G}_{\tau_l}}  & 
\alpha(\demo{act(a,o)}) :={\mathcal{A}_{a,\tau_o}} & 
\alpha(\demo{act(a,o,o')})  :={\mathcal{A}_{a,\tau_o,\tau_o'}}
\end{array}
\]
where $\tau_l, \tau_o$ denote the type of location $l$ and object $o$ respectively. In other words, when abstracting a trace as a string, the algorithm replaces specific object instances with their corresponding types. Intuitively, this abstraction captures the commonality between different actions in the trace, allowing generalization from a specific sequence of actions to a more general program structure. 

\def\scalefactor{0.3 }
\begin{figure}[b]
\footnotesize
    \begin{mathpar}
  \begin{array}{lllll}
\rho_s & \in & \text{Item List} & := &  \stx{scanObj}(??_{\tau_o}) \ALT \stx{scanLoc}(??_{\tau_l}) 
\\
s & \in & \text{Sketch} & := &  
\stx{actUnary}(a,??_v:\tau_o) 
\ALT \stx{actBinary}(a,??_v:{\tau_o},??_v:{\tau'_o}) \ALT
\stx{goto}(??_{v})   \ALT 
\stx{if}(??_\phi) \{s\}    \ALT
\\
& & & &
\stx{foreach}(v \in \rho_s) \{s\} 
  \ALT
\stx{let} \;v:=\stx{getNth}(\rho_s,??_n)
\ALT
s;s
    
\end{array}
  \end{mathpar}
  \vspace{-0.08in}
  \caption{Syntax of Program Sketches. Domain specific definitions (i.e., $\tau_o, \tau_l, a$)  
 are identical to Figure~\ref{fig:syntax}.}
  \label{fig:sketch_syntax}
  \vspace{-0.05in}
\end{figure}

Next, given the string abstraction $A$ of the demonstrations $\mathcal{D}$,  the loop in lines 3--12 alternates between the following  key steps:
\begin{itemize}[leftmargin=*]
     \item {\bf Regex synthesis:} The \textsf{GetNextRegex} procedure at line 4 finds a regular expression $r$ matching all strings in $A$.  Intuitively, this regex captures the main control flow structure of the target program and can be used to generate a set of program sketches.
     \item {\bf Sketch generation:} The inner loop in lines 6--10 translates a given regex to a \emph{set} of program sketches. As shown in Figure~\ref{fig:sketch_syntax}, a sketch has almost the same syntax as programs in our DSL except that the arguments of most constructs are unknown, as indicated by question marks. In particular, note that (1) the types of objects and locations being scanned are unknown, (2) predicates of $\stx{if}$ statements are yet to be determined, and (3) the specific objects and locations being acted on are also unknown (although their \emph{types} are known).  
     \item {\bf Sketch completion:} Given a candidate program sketch $s$, line 9 of the algorithm invokes \textsf{CompleteSketch} to find a completion $\pi$ of $s$ that is consistent with the demonstrations. If \textsf{CompleteSketch} does not return failure ($\bot$), the synthesized policy is guaranteed to satisfy all demonstrations; hence, {\sc Synthesize} returns $\pi$ as a solution at line 10.
     \end{itemize}

In the remainder of this section, we describe sketch generation and sketch completion in more detail. Because learning regexes from a set of positive string examples is a well-understood problem, we do not describe it in this paper, and our implementation uses an off-the-shelf tool customized to our needs via some post-processing (see {Section~\ref{sec:impl}}).

\subsection{Sketch Inference}

Given a regex $r$ over the alphabet $\Sigma = \{ \mathcal{G}_{\tau_l}, \mathcal{A}_{a, \tau_o}, \mathcal{A}_{a, \tau_o, \tau'_o} \}$, the goal of sketch inference is to (lazily) generate a set of program sketches. The inputs to the sketch inference procedure are regular expression of the following form:
\[
  \begin{array}{lcl}
{r}  & := & 
 {\mathcal{A}_{a,\tau_o}}
\ALT 
 {\mathcal{A}_{a,\tau_o, \tau_o'}}
\ALT 
 {\mathcal{G}_{\tau_l}}
\ALT 
  {rr}
 \ALT 
 {(r)^*} 
 \ALT 
 {(r)?} 
 \end{array}
 \]
Given such a regex, sketch inference consists of two steps:

\begin{enumerate}[leftmargin=*]
    \item {\bf Syntax-directed translation:} In the first step, sketch inference converts the given regex to control flow operations using  syntax-directed translation. Intuitively, string concatenation is translated into to sequential composition; Kleene star corresponds to loops; and, optional regexes translate into conditionals. 
    \item {\bf Perception inference:} While the sketches generated in step (1) are syntactically valid, they may lack essential perception operations (i.e., $\stx{scanObj}$ and $\stx{scanLoc}$). 
    Hence, in the second step, our sketch inference procedure inserts these perception operations such that the resulting sketch is \emph{perception-complete}, meaning that it contains at least the minimum number of required scan operation. However, since the target program may require additional $\stx{scan}$ operations, the second step of sketch inference yields a set of sketches that only differ with respect to the placement of these perception operations. 
    \end{enumerate}

\def\horizontalSpace{2mm}
\begin{figure}[t]
\footnotesize
\[
\RULE{}{{\mathcal{G}_{\tau_l}} \rhd \stx{goto}(??_v:\tau_l)}
\qquad 
\RULE{}{{\mathcal{A}_{a,\tau_o}} \rhd \stx{actUnary}(a, ??_v:\tau_o)}
\qquad
\RULE{}{{\mathcal{A}_{a,\tau_o, \tau'_o}} \rhd \stx{actBinary}(a, ??_v:\tau_o, ??_v:\tau'_o)}
\]\\
\[
\RULE{r\rhd s}{(r)^*\rhd \stx{foreach}(v \in \stx{scanLoc}(??_{\tau_l}))\{s\}}
\qquad
\RULE{r\rhd s}{(r)^*\rhd \stx{foreach}(v \in \stx{scanObj}(??_{\tau_o}))\{s\}}
\qquad 
\RULE{r_1\rhd s_1\quad r_2\rhd s_2}{r_1 r_2 \rhd s_1;s_2}
\]\\
\[
\RULE{r\rhd s}{ (r)? \rhd \stx{if}(??_\phi)\{s\}}
\qquad
\frac{r\rhd s}{r\rhd \stx{let}\;v:= \stx{getNth}(\stx{scanObj}(??_{\tau_o}),??_n);s}
       \qquad
    \frac{r\rhd s}{r\rhd \stx{let}\;v:= \stx{getNth}(\stx{scanLoc}(??_{\tau_l}),??_n);s}
\]
\vspace{-0.03in}
\caption{Regex to sketch inference rules.  
}
\label{fig:regex_to_sketch}
\vspace{-0.03in}
\end{figure}

    Figure~\ref{fig:regex_to_sketch} presents our syntax-directed translation rules for converting regular expressions to a syntactically valid sketch using judgments of the form $r \triangleright s$, meaning that regex $r$ is translated to sketch $s$. As expected, characters $\mathcal{G}_\tau, \mathcal{A}_{a, \tau}, \mathcal{A}_{a, \tau, \tau'} $ are translated to $\stx{goto}$, $\stx{actUnary}$, and $\stx{actBinary}$ constructs respectively.  
    The Kleene star operator is translated into a looping construct, but may iterate either over locations or objects. Finally, regex concatanation is translated into sequential composition, and $(r)?$ is translated into a conditional with an unknown predicate.

Recall that our DSL also allows $\stx{let}$ bindings that assign a new variable to the result of a perception operation. Since program traces (and, hence, the inferred regexes) do not contain these perception operations, the last two rules in Figure~\ref{fig:regex_to_sketch} 
 allow inserting $\stx{let}$ bindings at arbitrary positions in the sketch. In particular, if $r$ can be translated into a sketch $s$, then the last two rules of Figure~\ref{fig:regex_to_sketch} state that $r$ can also be translated into a sketch of the form $l; s$ where $l$ is a new $\stx{let}$ binding which assigns a fresh variable $v$ to an entity that is obtained by scanning objects or locations. 

In general, observe that a  regex can give rise to a large number of program sketches, as we do not a priori know where to insert $\stx{let}$ bindings. To tackle this problem, our lazy sketch inference procedure first translates a regex into a sketch \emph{without} using the last two rules in Figure~\ref{fig:regex_to_sketch} for inserting $\stx{let}$ bindings. In a second step, it infers where perception operations are needed and inserts let bindings according to the results of this analysis. 

This second step of our sketch inference procedure is formalized using the notion of \emph{perception completeness}. Intuitively, a sketch is \emph{perception complete} if the program perceives (using $\stx{scan}$ operations) all objects that it manipulates, before it manipulates them. If a sketch is \emph{not} perception complete, it can never be realized into a valid program; hence, it is wasteful to consider such sketches. We  formalize the notion of perception completeness using the following definition:

 \begin{definition}{\bf (Perception Completeness)}
 Let $\prec$ denote a standard partial order between program points.\footnote{We refer the interested reader to the chapter 6 of "Programming Language Pragmatics" by~\citet{scott2000programming} for a detailed discussion on program orders and their usage in control flow analysis.}
 A sketch $s$ is said to be  \emph{perception complete} if the following conditions are satisfied:
 \begin{enumerate}[leftmargin=*]
     \item for all  $s_1:=\stx{actUnary}(a, ??:\tau_o)$ in $s$, there exists a  $s_2:=\stx{scanObj}(\tau_o)$ such that $s_2\prec s_1$.
     \item for all  $s_1:=\stx{actBinary}(a, ??:\tau_o, ??:\tau'_o)$ in $s$, there exist  $s_2:=\stx{scanObj}(\tau_o)$ and $s_3:=\stx{scanObj}(\tau'_o)$, such that $s_2\prec s_1$ and $s_3\prec s_1$.
     \item for all  $s_1:=\stx{goto}(??:\tau_l)$ in $s$, there exists a statement $s_2:=\stx{scanLoc}(\tau_l)$ in $s$ such that $s_2\prec s_1$.
 \end{enumerate}
\end{definition}

Our sketch inference algorithm leverages this notion of perception completeness to lazily enumerate program sketches as follows: First, it translates a given regex into a set of sketches using the inference rules shown in Figure~\ref{fig:regex_to_sketch} but \emph{without} using the last two  rules. It then infers a minimal set of applications of the last two rules needed to make the sketch perception complete and then augments the resulting sketches with the inferred $\stx{let}$ bindings. Finally, because additional $\stx{let}$ bindings may be needed, it lazily inserts more $\stx{let}$ bindings (up to a bound) if the current sketch does not produce a valid completion.

\subsection{Sketch Completion}

{
\begin{algorithm}[!t]
\scriptsize
\caption{Sketch Completion Algorithm }\label{alg:comp}
\KwInput{A set of demonstrations $\mathcal{D}$, a sketch $s$}\\
\KwOutput{A completion of $s$ consistent with $\mathcal{D}$ or $\bot$ if none exists}\\
\begin{algorithmic}[1]
\STATE $\mathsf{CompleteSketch}_\theta(s, \mathcal{D})\{$
\STATE \ \ \ \ $W:=[(s, {1.0})]$ \ \ \ \algcmt{the given sketch $s$ is the only partial program initially}
\STATE \ \ \ \ {\bf while} ($W\not\eq\emptyset$) \{
\STATE \ \ \ \ \ \ \ \ $(\partial,p) := W.\mathsf{dequeue}()$ \ \ \ \algcmt{get the partial program with the highest probability}
\STATE \ \ \ \ \ \ \ \ {\bf if} $(\mathsf{IsComplete}(\partial))\{$ \ \ \ \algcmt{check if there is any holes left to be filled}
\STATE \ \ \ \ \ \ \ \ \ \ \ \ {\bf if} $(\partial \models \mathcal{D})$ {\bf return }$\partial$ \ \ \ \algcmt{return the completed program if it is consistent with the demos}
\STATE \ \ \ \ \ \ \ \ \ \ \ \ {\bf else} {\bf continue}
\STATE \ \ \ \ \ \ \ \ {\bf if} $\neg \mathsf{Compatible}(\partial, \mathcal{D})\;{\bf continue}$ \ \ \ \algcmt{check compatibility of the partial program and demos}
\STATE \ \ \ \ \ \ \ \ $h := \mathsf{GetNextHole}(\partial)$
\STATE \ \ \ \ \ \ \ \ {\bf foreach} $(c \in \mathsf{Fill}(\partial,h, \mathcal{D}))\{$
\STATE \ \ \ \ \ \ \ \ \ \ \ \ $\partial' := \partial[h\mapsto c]$; $p' := \theta(c \ | \ \partial, h, \demos)$ \ \ \ \algcmt{fill the chosen hole and assign a probability to the result}
\STATE \ \ \ \ \ \ \ \ \ \ \ \ $W.\mathsf{enqueue}((\partial',p'))$ \ \ \ \algcmt{add the new partial program to the worklist}
\STATE \ \ \ \ \ \ \ \ $\}\}\}$
\end{algorithmic}
\end{algorithm}
}

We now turn our attention to the sketch completion procedure, shown in Algorithm~\ref{alg:comp}, for finding a sketch instantiation that satisfies the given demonstrations.  Given a sketch $s$ and demonstrations $\mathcal{D}$, \textsf{CompleteSketch} either returns $\bot$ to indicate failure in finding a policy $\pi$ that is consistent with all demonstrations. Note that the sketch completion procedure is parameterized over a statistical model $\theta$ for assigning probabilities to possible sketch completions.

 \textsf{CompleteSketch} is a  standard top-down  search procedure that iteratively expands partial programs until a solution is found.\footnote{Partial programs belong to the grammar from  Figure~\ref{fig:syntax} augmented with  productions  $M \rightarrow ??$ for each non-terminal $M$.} However, our sketch completion procedure has two novel aspects: First, it assigns probabilities to partial programs using a large language model, and second, it uses a novel \emph{compatibility checking} procedure for proving unrealizability of synthesis problems.

In more detail, the sketch completion procedure initializes the worklist to a singleton containing the input sketch $s$, with corresponding probability $1.0$ (line 2). It then enters a loop (lines 3--14) where each iteration processes the highest probability partial program $\partial$ in the worklist. If the dequeued partial program $\partial$ is complete, meaning that it has no holes (line 5), the algorithm checks whether all demonstrations are satisfied (line 6). If so, $\partial$ is returned as a solution; otherwise, it is discarded. If $\partial$ is incomplete, the algorithm performs a \emph{compatibility} check at line 8 to avoid solving an unrealizable synthesis problem.  Next, if $\partial$ is a compatible with the demonstrations, the algorithm considers one of the holes $h$ in $\partial$ and \emph{all} well-typed grammar productions that can be used to fill that hole. In particular, given a hole $?_N$, the procedure \textsf{Fill} yields a set of 
expressions $c_1, \ldots, c_n$ such that (1) $N \rightarrow c_i$ is a production in the grammar, and (2) replacing $h$ with $c_i$ can result in a well-typed program. Hence, for each such expression $c_i$, we obtain a new partial program $\partial'$ at line 11 by replacing hole $h$ in $\partial$ with expressions $c_i$.\footnote{When doing the replacement $\partial[h \mapsto c]$, all non-terminals $N$ occurring in $c$ are replaced with a hole $?_N$.}   
However, since some completions are much more likely than others, our algorithm assigns probabilities to completions using the statistical model $\theta$. Hence, when dequeuing partial programs from the worklist, the algorithm prioritizes programs that are assigned a higher probability. 




\subsubsection{Proving Unrealizability}\label{sec:feasible}

{
\begin{algorithm}[!t]
\scriptsize
\caption{Checking compatibility between partial programs and demonstrations}\label{alg:feas}
\KwInput{A partial program $\partial$, a set of demonstrations $\mathcal{D}$}\\
\KwOutput{True if the given partial program is compatible with $\mathcal{D}$ and otherwise false}\\
\begin{algorithmic}[1]
\STATE $\mathsf{Compatible}(\partial, \mathcal{D})$
\STATE \ \ \ \ {\bf foreach } $((\env,t)\in\mathcal{D})\; $
\STATE \ \ \ \ \ \ \ \ $\partial^* := \mathsf{PartialEval}(\partial, \env)$ \algcmt{Partially evaluate the given partial program on the demonstration environment}
\STATE \ \ \ \ \ \ \ \ $r := \mathsf{ProgToRegex}(\partial^*, \alpha(\env))$ \algcmt{Find the regex that over-approximates behaviors of $\partial^*$}
\STATE \ \ \ \ \ \ \ \  {\bf if}$(\alpha(t) \not \in r)\;$   {\bf return} $\emph{false}$ \algcmt{check if the trace is not accepted by the over-approximating regex}
\STATE \ \ \ \ {\bf return} $true$
\end{algorithmic}
\end{algorithm}
}

{A key component of our sketch completion procedure is a novel technique for proving unrealizability of a programming-by-demonstration (PbD) problem. While there has been significant prior work on proving unrealizability in the context of programming-by-example (PbE)~\cite{10.1007/978-3-030-25540-4_18, 10.1145/3571216,FM+18}, such techniques only consider the input-output behavior rather than the entire execution trace. In contrast, our goal is to prove unrealizability of synthesis problems where the specification is a set of demonstrations (i.e., traces).}

Given a partial program $\partial$ (representing a hypothesis space), our key idea is to generate a regex $r$ such that the language of $r$ includes all possible traces of all programs in the hypothesis space. Hence, if there exists some trace $t \in \demos$ where $\alpha(t)$ is not accepted by $r$, this constitutes a proof that the synthesis problem $(\partial, \demos)$ is unrealizable. 


Our algorithm for checking compatibility between partial programs and traces is presented in Algorithm~\ref{alg:feas}.
At a high level, this algorithm iterates over all demonstrations (lines 2--5) and returns false if it can prove that $\partial$ is incompatible with some demonstration $(\mathcal{E}, t)$ in $\mathcal{D}$. To check compatibility with $(\mathcal{E}, t)$, the algorithm first partially evaluates $\partial$ on the initial environment $\mathcal{E}$ to obtain a simplified program $\partial^*$, as done in existing work~\cite{morpheus}. The novel part of our technique lies in constructing a regex abstraction of the partial program $\partial$ under a given environment $\mathcal{E}$. Specifically, our compatibility checking procedure constructs a  regex $r$ that  \textit{over-approximates} the possible behaviors of $\partial$ under initial environment $\mathcal{E}$. In particular, the regex $r$   is constructed at line 4 in such a way that if $\alpha(t)$ is not accepted by $r$, then no completion of $\partial$ can be compatible with $(\mathcal{E}, t)$ (see, Theorem~\ref{th:completeness}).

Hence, the crux of the compatibility checking algorithm is the \textsf{ProgToRegex} procedure (formalized as  inference rules shown in Figures~\ref{fig:over_apprx_scan_rules}~and~\ref{fig:partial_to_regex}) for generating a regex that over-approximates the behavior of $\partial$ under environment $\mathcal{E}$. 
These rules utilize the notion of an \emph{abstract environment} which is a triple  $\absenv:=(\mathsf{CurLoc}, \mathsf{Locs}, \mathsf{Objs})$ where (1)  \textsf{CurLoc} is a set containing all possible locations that the  robot \emph{could} currently be at; (2) \textsf{Locs} is a mapping from location types to the set of locations of that type; (3) \textsf{Objs} is a mapping from each location to the set of objects of each type at that location (or $\top$ if unknown).
Because statements in our DSL can modify the environment, this notion of abstract environment is used to conservatively capture (the possibly unknown) side effects of partial programs on the environment. 
The inference rules shown in Figures~\ref{fig:over_apprx_scan_rules}~and~\ref{fig:partial_to_regex} formalize the \textsf{ProgToRegex} procedure using two types of judgments:
\begin{enumerate}[leftmargin=*]
    \item {\bf Scan rules} (shown in Figure~\ref{fig:over_apprx_scan_rules}) are of the form 
    $\absenv \vdash \stx{scan}(...): \Theta$,  indicating that the \emph{cardinality} of the set returned by $\stx{scan}$ must be some element of $\Theta$. For example, if $\Theta= 
    \{1, 4\}$, this means that the number of objects/locations returned by scan is either $1$ or $4$. On the other hand, if $\Theta$ includes the special $\star$ element, then the number of elements is unknown. 
    \item {\bf Partial program rules} (shown in Figure~\ref{fig:partial_to_regex}) are of the form $\absenv \vdash \pi: \absenv', r$, meaning that, under initial abstract environment $\absenv$,  (a) the behavior of $\pi$ is over-approximated by regex $r$, and (b) $\absenv'$ is a new  environment that captures all possible environment states after executing $\pi$ on $\absenv$.
\end{enumerate}

Before we explain  these rules in detail, we first describe the high level idea, which  is to encode (a)  atomic actions using characters drawn from the alphabet $\{ \mathcal{G}_{\tau}, \mathcal{A}_{a, \tau}, \mathcal{A}_{a, \tau, \tau'} \}$, (b) $\stx{if}$ statements using optional regexes, and (c) loops using regexes of the form $r^n$ where $n$ denotes the number of times the loop will execute (or as $r^*$ if the number of loop iterations is completely unknown). {As mentioned in earlier sections (and, as we demonstrate empirically in Section~\ref{sec:eval}), static reasoning about the number of loop iterations improves the effectiveness of our unrealizability checking procedure.}
With this intuition in mind, we now explain the rules shown in Figures~\ref{fig:over_apprx_scan_rules}~and~\ref{fig:partial_to_regex}. 

\paragraph{Scan rules.} There are two sets of rules for scan operations, (\textsc{loc-known}) and (\textsc{loc-unknown}) for scanning locations, and (\textsc{obj-known}) and (\textsc{obj-unknown}) for scanning objects. For a $\stx{scanLoc}$ operation, if its argument is a known location type $\tau_l$, we simply look up the number of locations of that type from the given abstract environment. If it is unknown, we take the union over all possible location types. The rules for $\stx{scanObj}$ are similar: 
The (\textsc{Obj-Known}) rule handles the case where the argument is a known object type $\tau_o$. In this case, we consider all the locations that the robot could be currently at and take the union of the number of objects of type $\tau_o$ for all of those locations. In the \textsc{Obj-Unknown} rule, we additionally take the union over all possible object types, since the argument of the $\stx{scanObj}$ operation is unknown. 

\paragraph{Atomic actions.} The rule labeled ({\sc atomic}) in Figure~\ref{fig:partial_to_regex} deals with $\stx{goto}, \stx{actUnary}$, and $\stx{actBinary}$ statements and  serves two roles. First, it abstracts the performed action as a letter in our regex alphabet using the abstraction function $\alpha$. Second, it produces a new abstract environment $\absenv'$ by considering all possible effects of the action on the input environment via the \textsf{UpdateAbsEnv} function. Since the \textsf{UpdateAbsEnv} function is domain-specific and depends on the types of actions of interest, we do not describe in detail but provide a set of representative examples in Appendix~\ref{subapp:updateAbsEnv}.

\paragraph{Sequence.} Sequential composition is abstracted using  regex concatanation, and its final effect on the environment is captured by threading the environment through the two premises of the rule.

\paragraph{Conditionals.} As expected, if statements are abstracted using the optional operator ($(r)?$). Furthermore, since we do not know whether the predicate evaluates to true or not\footnote{Recall that we apply partial evaluation before performing this step. Hence, if the predicate of a conditional can be fully evaluated, it will be simplified away and rewritten as conditional-free code.}, we take the \emph{join} of the two abstract environments. Intuitively, the join of abstract environments $\absenv$ and $\absenv'$, denoted by $\absenv\sqcup\absenv'$, is the smallest environment that over-approximates both $\absenv$ and $\absenv'$. An abstract environment $\absenv$ is said to over-approximate an abstract environment $\absenv'$, denoted by $\absenv'\sqsubseteq\absenv$, if and only if: 
\[\absenv'.\mathsf{Locs}\subseteq \absenv.\mathsf{Locs} \wedge \forall_{l\in\absenv'.\mathsf{Locs}}\forall_{\tau\in\mathsf{ObjTypes}(\absenv')}.\; \absenv'.\mathsf{Objs}(l,\tau) \subseteq \absenv.\mathsf{Objs}(l,\tau) \]
Then, we can define a join operator on abstract environments as follows:
\[
\absenv\sqcup\absenv' = \absenv'' 
\iff
(\absenv\sqsubseteq\absenv'')
\wedge
(\absenv'\sqsubseteq\absenv'')
\wedge 
(\forall_{\absenv'''} (\absenv\sqsubseteq\absenv''' 
\wedge
\absenv'\sqsubseteq\absenv''' \Rightarrow \absenv''\sqsubseteq\absenv'''))
\]
Intuitively,  $\absenv'''$ is the result of joining $\absenv$ and $\absenv'$ if it is the smallest  abstract environment that over-approximates both.

\def\horizontalSpace{1mm}


\begin{figure}[t]
\vspace{-0.2in}
\footnotesize
\begin{minipage}[t]{0.32\textwidth}
  \ruleLabel{loc-known}
$$
\RULE{
\\
}{
\absenv\vdash \stx{scanLoc}(\tau_l):\{|\absenv.\mathsf{Locs}(\tau_l)| \}
}$$
\end{minipage}
\qquad
\begin{minipage}[t]{0.38\textwidth}
  \ruleLabel{loc-unknown}
$$
\RULE{
\mathsf{LocTypes}(\absenv) = \{\tau_1,\dots,\tau_n\}
}{
\absenv\vdash \stx{scanLoc}(??):\bigcup_{1\leq i\leq n}\{|\absenv.\mathsf{Locs}(\tau_i) |\}
}$$
\end{minipage}
\\[\horizontalSpace]
\begin{minipage}[t]{0.41\textwidth}
  \ruleLabel{obj-known}
$$
\RULE{
\absenv.\mathsf{CurLocs} = \{l_1,\dots,l_n\}
}{
\absenv\vdash \stx{scanObj}(\tau_o):\bigcup_{1\leq i \leq n} \{|\absenv.\mathsf{Objs}(l_i,\tau_o)|\}
}$$
\end{minipage}
\hfill
\begin{minipage}[t]{0.54\textwidth}
  \ruleLabel{obj-unknown}
$$
\RULE{
\absenv.\mathsf{CurLocs} = \{l_1,\dots,l_n\}
\qquad 
\mathsf{ObjTypes}(\absenv) = \{\tau_{0},\dots,\tau_k \}
}{
\absenv \vdash \stx{scanObj}(??):\bigcup_{1\leq i\leq n}\bigcup_{1\leq j\leq k}\{|\env.\mathsf{Objs}(l_i,\tau_j)|\}
}$$
\end{minipage}
\caption{Over-approximation of robot perception used in \textsf{ProgToRegex} function. 
}
\label{fig:over_apprx_scan_rules}
\end{figure}

\begin{figure}[t]
\vspace{-0.1in}
\footnotesize
\begin{minipage}[t]{0.45\textwidth}
  \ruleLabel{atomic}
$$
\RULE{
\mathsf{AtomicAction}(s)
\qquad \absenv' = \mathsf{UpdateAbsEnv}(\absenv,s)
}{
\absenv\vdash s:\absenv',\alpha(s)
}$$
\end{minipage}
\qquad
\begin{minipage}[t]{0.33\textwidth}
  \ruleLabel{seq}
$$
\RULE{
\absenv\vdash s_1:\absenv_1,r_1
\qquad
\absenv_1\vdash s_2:\absenv_2,r_2
}{
\absenv \vdash s_1;s_2 : \absenv_2, r_1r_2
}$$
\end{minipage}
\\[\horizontalSpace]
\begin{minipage}[t]{0.28\textwidth}
  \ruleLabel{if}
$$
\RULE{
\absenv \vdash s:\absenv',r
}{
\absenv\vdash \stx{if}(\_)\,\{s\}: \absenv\sqcup\absenv',(r)?
}$$
\end{minipage}
\hfill
\begin{minipage}[t]{0.21\textwidth}
  \ruleLabel{let}
$$
\RULE{
\\
}{
\absenv \vdash \stx{let}\; v:= e :\absenv,\epsilon
}$$
\end{minipage}
\hfill
\begin{minipage}[t]{0.48\textwidth}
  \ruleLabel{loop}
$$
\RULE{
\absenv\vdash \rho_s:\{n_1,\dots,n_k\}
\qquad \absenv'\vdash s:\absenv',r
\qquad \absenv\sqsubseteq \absenv'
}{
\absenv\vdash \stx{foreach}(v\in\rho_s)\{s\}: \absenv', r^{n_1}|r^{n_2}|\dots| r^{n_k}
}$$
\end{minipage}
\caption{
Over-approximation of partial programs used in \textsf{ProgToRegex} function. 
}
\label{fig:partial_to_regex}
\end{figure}

\paragraph{Loops.} The last rule in Figure~\ref{fig:partial_to_regex} summarizes the analysis of loops. First, since the environment may be modified in the loop body, this rule first computes an \emph{inductive abstract environment}, $\absenv'$, for the loop ({see Section~\ref{sec:impl} for further details}). In particular, the premise $\absenv \sqsubseteq \absenv'$ ensures that $\absenv'$ over-approximates the \emph{initial} environment, thereby establishing our base case. Second, the premise $\absenv' \vdash \pi: \absenv', r$ ensures that $\absenv'$ is preserved in all iterations of the loop body. Furthermore, because $\absenv'$ is an over-approximation of the environment that the loop body $\pi$ operates in, the regular expression $r$ also over-approximares the behavior of $\pi$. Finally, to over-approximate the behavior of the entire loop, we determine the possible number of loop executions using the rules from Figure~\ref{fig:over_apprx_scan_rules} under the {initial} environment $\absenv$. If $\rho_s$ can yield $n$ different objects, then the behavior of the loop is captured as $r^n$. However, since we may not be able to compute the exact number of objects returned by a scan operation, we consider all possible cardinalities $n_1, \ldots, n_k$ of the resulting set. Hence, the behavior of the loop is captured by the disjunction of the regexes $r^{n_1}, \ldots, r^{n_k}$. 

{
The following theorem states that none of the completions of an unrealizable partial program is consistent with all demonstrations (see Appendix~\ref{subapp:proof} for the proof). An immediate corollary is the \textit{bounded completeness} of our search algorithm, i.e., that Algorithm~\ref{alg:top} always finds a program consistent with the demonstrations if it exists within the bounds of the search space. 
}

\begin{theorem}
Let $\partial$ be a partial program and let $\demos$  be a set of demonstrations. Then, for any complete program $P$ that is a completion of $\partial$, we have:
\[
P \models \demos \ \Longrightarrow\  \mathsf{Compatible}(\partial, \demos)
\]

\label{th:completeness}
\end{theorem}

\vspace{-0.08in}
\subsubsection{Using LLM for Sketch Completion}\label{sec:feasible}
\new{We conclude this section with a discussion of how \tool\ infers a probability distribution over partial programs by prompting a large language model.
Our prompting approach is inspired by the success of LLMs in addressing the ``\textit{Fill in the Middle (FIM)}'' challenges in NLP literature~\cite{fim}. We reduce the sketch completion task to a FIM problem, as described below. }

\new{
Given a partial program $\partial$ with a hole $h$ to fill, our approach encodes the \textit{context} of $h$ in $\partial$ as a natural language prompt with unknown \emph{masks}~\cite{bert}.
The prompt includes the set of valid completions for each mask, chosen by the $\mathsf{Fill}(.)$ procedure in Algorithm~\ref{alg:comp}, to ensure the resulting program is well-typed.
The LLM is then instructed to infer a probability distribution over the set of completions (represented by $\theta$ in Algorithm~\ref{alg:comp}).
This procedure is repeated whenever a partial program is expanded into a set of partial programs, to prioritize the enumerative search 
towards the intended program. 
}

To illustrate how \tool\ prompts the LLM, Figure~\ref{subfig:partial_program} (top) shows a partial program containing six unfilled holes, denoted as $??_i$. 
Figure~\ref{subfig:llm_prompt} (top) shows the prompt used for completing hole $??_1$, which essentially corresponds to a natural description of the program.\footnote{Note that the demonstration is implicitly encoded as part of the partial program  using  type information for each hole.} To generate such a prompt, our approach translates control-flow constructs to natural language in a syntax-directed way and replaces some of the holes ($??_1$ and $??_2$ in this example) with masks. Because the remaining holes $??_3 - ??_6$ will be replaced with synthetic variable names like $v_1, v_2$, they are not meaningful to the LLM; so our approach simply uses the types of these holes rather than masks when translating the partial program to natural language. As we can see from the LLM output in Figure~\ref{subfig:llm_prompt}, the highest likelihood completion of $??_1$ is deemed to be \emph{bed} by the model, so the sketch completion algorithm prioritizes this completion over other alternatives such as   \emph{chair} or \emph{mug}.

 \algsetup{
    linenosize=\Small,
    linenodelimiter = {:}
}

\begin{figure}[h]

\scriptsize
 \begin{subfigure}[b]{0.49\textwidth}
 \centering
\begin{mdframed}[backgroundcolor=grey12, linecolor=black,innerleftmargin=0mm]
  \begin{algorithmic}[1]
  {
  \STATE $\stx{foreach}(v_1\in \stx{scanLoc}(room))\{$
  \STATE  \ \ \  $\stx{goto}(v_1:room)$;
   \STATE  \ \ \ $\stx{let}\; v_2:= \stx{getNth}(\stx{scanObj}(bin),0)$;
   \STATE  \ \ \  $\stx{actUnary}(open, v_2:bin);$
  \STATE  \ \ \ $\stx{foreach}(v_3 \in \stx{scanObj}(??_1))\{$
  \STATE  \ \ \   \ \   $\stx{foreach}(v_4 \in \stx{scanObj}(sheet))\{$
  \STATE  \ \ \  \ \ \  \ \ \   $\stx{if}(??_2)\{$
    \STATE  \ \ \  \ \ \  \ \ \  \ \ \    $\stx{actUnary}(grab,??_3:sheet);$
      \STATE  \ \ \  \ \ \  \ \ \  \ \ \    $\stx{actBinary}(put\text{-}in,??_4:sheet,??_5:bin);$
  \STATE  \ \ \  $\}$$\}$$\}$   
   \STATE  \ \ \   $\stx{actUnary}(close, ??_6:bin);\}$
  }
 \end{algorithmic}
 \end{mdframed}
 \vspace{1mm}
 \begin{mdframed}[backgroundcolor=grey12, linecolor=black,innerleftmargin=0mm]
  \begin{algorithmic}[1]
  {
\makeatletter
  \setcounter{ALC@line}{3}
\STATE \ \ \ \dots
  \STATE  \ \ \ $\stx{foreach}(v_3 \in \stx{scanObj}(bed))\{$
  \STATE  \ \ \   \ \   $\stx{foreach}(v_4 \in \stx{scanObj}(sheet))\{$
  \STATE  \ \ \  \ \ \  \ \ \   $\stx{if}(\stx{checkProp}(??_{2a},v_4)\wedge\stx{checkRel}(??_{2b},v_4,v_3))\{$
   \\ \ \ \ \dots
  }
 \end{algorithmic}
 \end{mdframed}
   \caption{Partial programs $\partial$ (top) and $\partial'$ (bottom)}
   \label{subfig:partial_program}
 \end{subfigure}
\hfill
\begin{subfigure}[b]{0.5\textwidth}
\begin{minipage}{\textwidth}
  \begin{mdframed}[linecolor=grey4,innerleftmargin=0.1cm]
  {
    \textbf{Prompt for $??_1$:}
    \prompt{For each room; Go to room; Look for bins; Open a bin;  For each $[M]_1$ do; For each sheet do; If $[M]_2$, grab sheet and put sheet in bin; Close bin.}\\[1mm]
    \textbf{LLM Output for $[M]_1$:}
    \{Bed: 0.8, Chair: 0.1, Mug: 0.05, \dots\}
}
 \end{mdframed}
 \end{minipage}
 \\[2.2mm]
 \begin{minipage}{\textwidth}

  \begin{mdframed}[linecolor=grey4,innerleftmargin=0.1cm]
  {
    \textbf{Prompt for $??_{2a}$:}
    \prompt{For each room; Go to room; Look for bins; Open a bin;  For each bed do; For each sheet do; If sheet is $[M]_{2a}$ and sheet is $[M]_{2b}$ bed, grab sheet and put sheet in bin; Close bin.}\\
    \textbf{LLM Output for $[M]_{2a}$:}
    \{Dirty: 0.75, Folded: 0.2, White: 0.02, \dots\}
}
 \end{mdframed}
 \end{minipage}
  \\[2.2mm]
\begin{minipage}{\textwidth}
  \begin{mdframed}[linecolor=grey4,innerleftmargin=0.1cm]
  {
    \textbf{Prompt for $??_{2b}$:}
    \prompt{For each room; Go to room; Look for bins; Open a bin; For each bed do; For each sheet do; If sheet is dirty and sheet is $[M]_{2b}$ bed, grab sheet and put sheet in bin. Close bin.}\\
    \textbf{LLM Output for $[M]_{2b}$:}
    \{On-top-of: 0.9, Under: 0.05, \dots\} 
}
 \end{mdframed}
 \end{minipage}
  \caption{LLM prompts for $??_1$, $??_{2a}$ and $??_{2b}$}
    \label{subfig:llm_prompt}
  \end{subfigure}
    \caption{Partial programs during synthesis and the generated LLM prompts to choose the next completion}
    \label{fig:llm_example}
\end{figure}


As another example, consider the process of filling hole $??_2$, and suppose that the algorithm has already refined $??_2$ to the conjunct
 $\stx{checkProp}(??_{2a},v_4)\wedge\stx{checkRel}(??_{2b},v_4,v_3)$
 as shown in the bottom part of Figure~\ref{subfig:partial_program}. When generating the prompt for $??_{2a}$, both holes $??_{2a}$ and $??_{2b}$ are filled with masks, and the LLM outputs \emph{dirty} as the most likely completion for $??_{2a}$. When querying the remaining hole ($??_{2b}$), $??_{2a}$ has already been filled with \emph{dirty}, so the prompt only contains a single mask, and the LLM outputs \emph{on-top-of} as the most likely completion.  As illustrated by these examples, the LLM-guided search strategy allows the sketch completion engine to quickly home in on the right concepts (such as \emph{bed}, \emph{dirty}, and \emph{on-top-of} in this example) and therefore  allows the search procedure to focus on the most promising sketch completions.

\section{Implementation}
We  implemented the proposed approach in a   tool called \tool\ written in Python.
%
 In this section, we discuss salient aspects of \tool\  that are not covered in the technical sections.

\mypar{Regex Learner}
Our implementation leverages an open-source library to learn regular expressions from positive samples~\cite{XML_Learner}. However, since our sketch generation procedure does not allow arbitrary regex operators, our implementation post-processes the synthesized regexes by applying a set of rewrite rules.
For instance, because our sketch generation procedure does not allow arbitrary disjunction, one of the rewrite rules ($(x|y) \rightarrow (xy?)$)  replaces  disjunction with an optional operator.
The full list of our rewrite rules can be found in Appendix~\ref{app:rewrite_rules}.
%

\mypar{Large Language Model}
Our sketch completion module utilizes the BERT large language model~\cite{bert}, with
pre-trained weights obtained from the HuggingFace library~\cite{wolf-etal-2020-transformers}. Since our algorithm generates \textit{masked language modeling} (MLM) queries, we chose to use the {\tt bert-base-uncased} model. This model is primarily fine-tuned for tasks that make use of the whole sentence, potentially with masked words, to make decisions. It is a lightweight model, and in our experiments, it consistently returns responses in less than $70\,ms$ on average.

\mypar{Parallel Sketch Completion}
The inherent independence of program sketches naturally lends itself to parallelization of the search process. To take advantage of this, \tool\ spawns a new process to execute sketch completion (Algorithm~\ref{alg:comp}) for each generated sketch.

\mypar{Computing Loop Invariants}{Recall that our procedure for proving unrealizability relies on an \emph{inductive abstract environment} for translating loops to regular expressions. Our implementation conservatively models the effect of any statement with holes by assuming that the hole could be filled with any value, and it performs standard least fixed point computation under these conservative semantics. }

\label{sec:impl}

\section{Evaluation}
\label{sec:eval}
In this section, we present experiments  designed to answer the following research questions:

\begin{enumerate}[leftmargin=3em]
    \item[{\bf (RQ1)}] How  effective is \tool \ at learning  policies from human demonstrations? 

\item[{\bf{(RQ2)}}] What is the relative significance of each of the key components in our synthesis algorithm?
    
    \item[{\bf (RQ3)}] How does \tool\ compare against relevant baselines 
    in terms of  learning  policies that match the user's intention?
    
  
\end{enumerate}

\newcommand{\prolexds}{\textsc{prolex-ds}}

\begin{figure}[t]
\centering
\footnotesize
    \begin{subfigure}[b]{0.58\textwidth}
     \centering
\includegraphics[height=29mm]{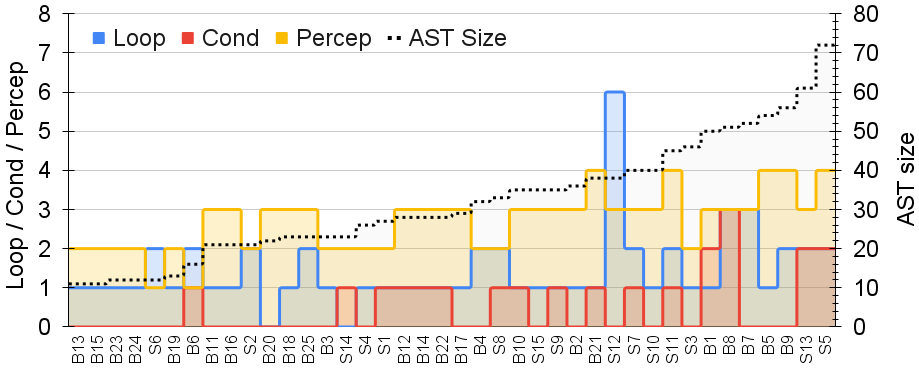}
 \caption{Complexity of the Tasks}
   \label{subfig:complexity_of_tasks}
\end{subfigure}
\hfill
 \begin{subfigure}[b]{0.4\textwidth}
 \centering
 \scriptsize
  \begin{tabular}{|c|c|c|c|}  
      \hline \centering
      \multirow{2}{*}{\textbf{Env}} & {\textbf{\# Obj.} } &{\textbf{\# Obj.}}   & \multirow{2}{*}{\textbf{\# Props.}}
        \\
        & \textbf{Types}& {\textbf{Instances}} &  
     \\
 \hline
    Easy &  40 & 140 & 609\\
    Medium & 60 & 295 & 2455\\
    Hard &  80 & 1109 & 13944 \\
 \hline
\end{tabular}
\vspace{3mm}
\caption{Complexity of the Environments}
\label{subfig:envs_summary}
\end{subfigure}
\caption{Overview of the Benchmark}
  \label{fig:env}
\end{figure}

\subsection{Benchmarks and Experimental Set-up}
\seclabel{bench}

\vspace{-0.05in}
\mypar{Tasks} 
\new{
We evaluate \tool\ on a set of 40 programmatic LfD problems involving long-horizon service tasks in typical household environments. %
We gathered these tasks from two sources, which we describe below.}

\new{
First, we use 25 tasks from the Behavior Project~\cite{pmlr-v164-srivastava22a}. This project is an interactive platform designed for a virtual embodied AI agent operating within simulated household environments. The Behavior Project offers simulation videos of the agent carrying out a range of long-horizon tasks. These demonstrations are produced by human users who use a joystick to control the agent's movements and actions in the simulation environment.
Out of 100 tasks provided in the Behavior project, we observed that 75 required precise low-level motion control of the robot's arms, like detailed cleaning of a car's surfaces. As our DSL can trivially execute these tasks without any conditionals or loops, we excluded them from our evaluation.
The demonstration videos for the remaining 25 tasks were transcribed using our action alphabet by an expert annotator ---one of the authors--- who is knowledgeable about the skill abstractions used by the robot.\footnote{\new{Note that, in principle, this step could be performed by an off-the-shelf vision-to-action detection (V2A) algorithm~\cite{ashutosh2023hiervl,nazarczuk2020v2a}. However, since this perception inference is orthogonal to our main problem, we opt to directly use the expert annotations.}}
}

\new{
Next, to broaden the scope of our evaluation and include more realistic tasks, we surveyed students at our institution and collected 15 additional tasks to evaluate \tool\ on. 
We did not impose any requirements regarding the users' technical expertise, as our aim was to gather a list of household tasks that a typical end-user would consider beneficial to have automated
}


%
%


%
%
A full list of the above 40 tasks and their ground-truth programs is provided in Appendix~\ref{app:bench_task_desc}.
{
Figure~\ref{subfig:complexity_of_tasks} presents an overview of the ground-truth programs for these tasks, ordered by  program size. 
In more detail, the right y-axis represents program size (measured in terms of the number of nodes in the abstract syntax tree) and the left y-axis represents the number of loops, conditionals and perception primitives in the program. Tasks from the Behavior project are labeled as $\mathtt{B1}\dots \mathtt{B25}$, and surveyed tasks are labeled as $\mathtt{S1}\dots \mathtt{S15}$.
}

\begin{wrapfigure}[11]{r}{0.35\textwidth}
  \centering
  \vspace{-2mm}
  \includegraphics[width=0.33\textwidth]{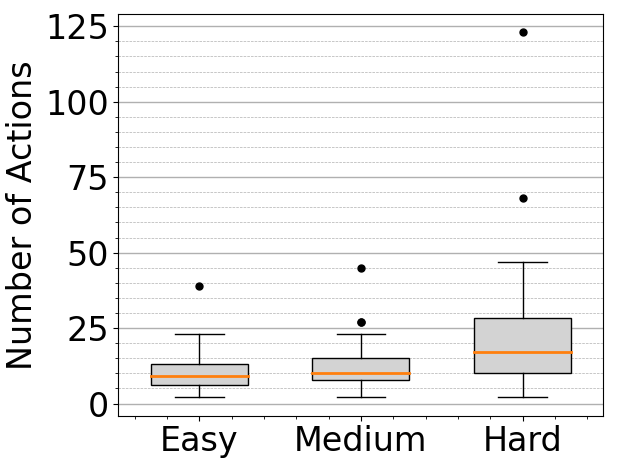}
  \caption{Length of Demonstrations}
  \label{fig:demo-len}
\end{wrapfigure}

\mypar{Environments} All of our benchmarks are defined in three  household environments,  summarized in  Figure~\ref{subfig:envs_summary}. Because the difficulty of the synthesis task depends crucially on the number of object types and objects in the environment, we classify the three environments as Easy, Medium, and Hard based on these numbers. 
As we can see from Figure~\ref{subfig:envs_summary}, these environments contain  up to thousands of objects and over ten thousand properties. 


\mypar{Full benchmark set} Overall, we evaluate \tool\ on a total of 120 benchmarks, with 40 unique tasks  and 3 different environments.  For each of the 40 tasks, we manually write the ground truth program in our DSL and obtain a demonstration by running the ground truth program. \new{Figure~\ref{fig:demo-len} presents statistics regarding the length of these demonstrations. On average, demonstrations in the easy, medium, and hard environments consist of 11, 13, and 24 actions, respectively. Some demonstrations in the hard environment exceed 100 actions due to the large number of object instances that need to be handled. }

\mypar{Experimental set-up} 
Our experiments were conducted on a 
server with 64 available Intel Xeon Gold 5218 CPUs @2.30GHz, 264GB of available memory, and running Ubuntu 22.04.2. We use a time limit of 120 seconds per task in all of our experiments.

\subsection{Main  Results for Prolex}
\begin{figure}[t]
 \begin{subfigure}[b]{0.45\textwidth}
  \centering
 \includegraphics[height=33mm]{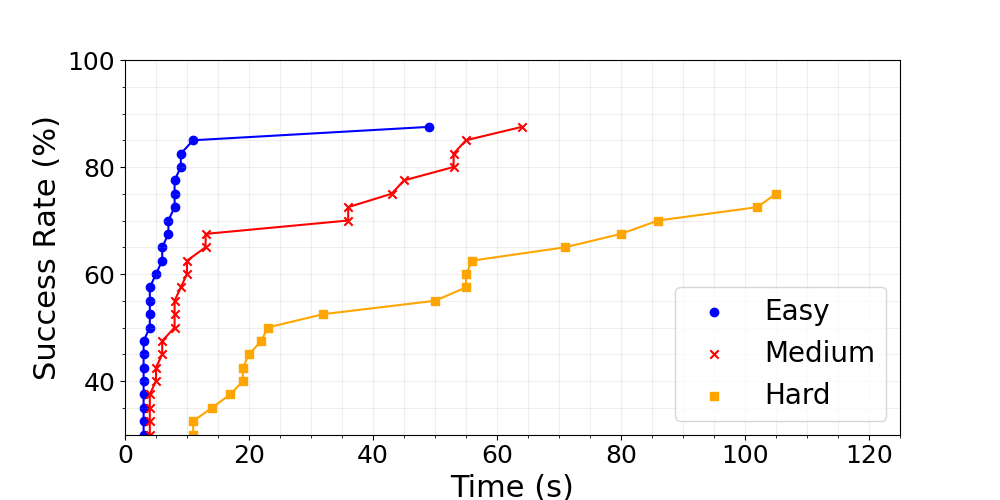}
  \caption{Success Rate per Environment}
    \label{subfig:succes_cactus}
\end{subfigure}
\hfill
 \begin{subfigure}[b]{0.26\textwidth}
  \centering
 \includegraphics[height=29mm]{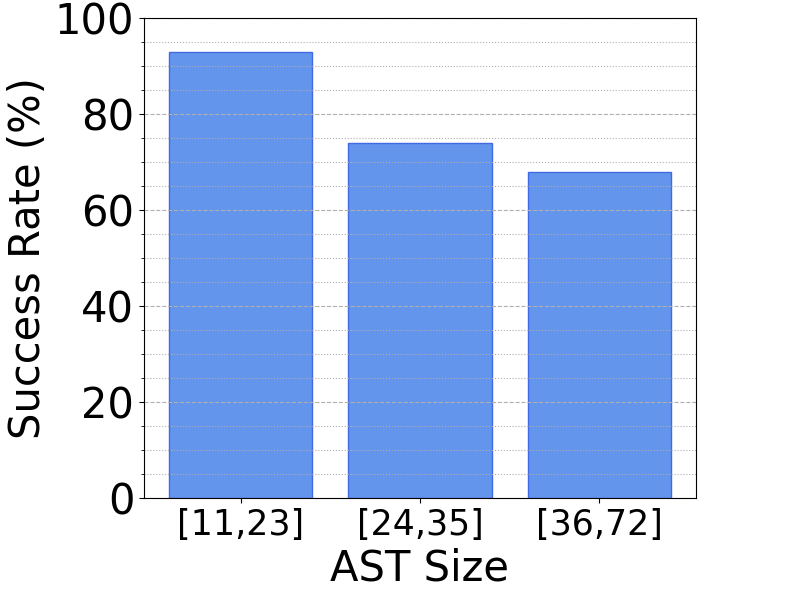}
  \caption{Success by AST Size}
    \label{subfig:succes_vs_ast}
\end{subfigure}
\hfill
 \begin{subfigure}[b]{0.26\textwidth}
  \centering
 \includegraphics[height=29mm]{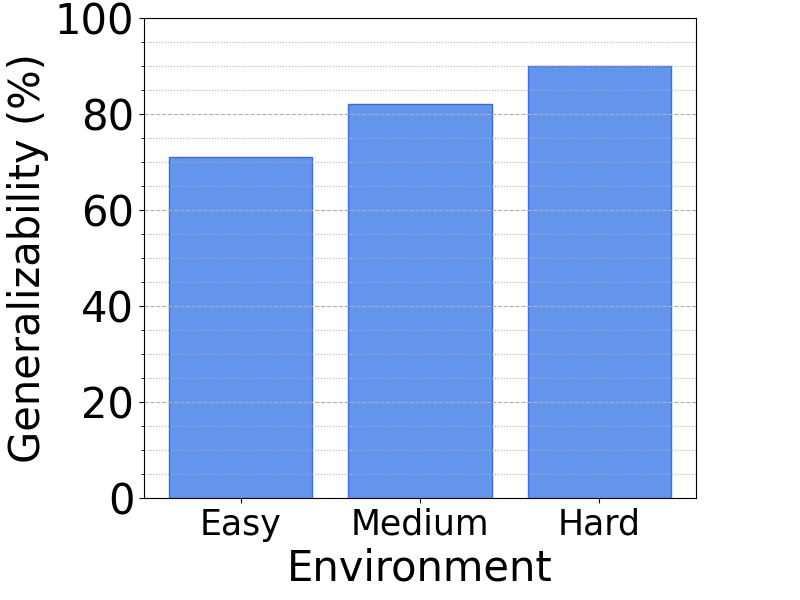}
  \caption{Generalizability}
    \label{subfig:gen_vs_env}
\end{subfigure}
    \caption{Experimental Results}
\label{fig:results}
\end{figure}

Our main results are presented in {Figure~\ref{fig:results}}, with Figure~\ref{subfig:succes_cactus} showing the percentage of completed tasks against synthesis time. We consider a task to be \emph{completed} if \tool\ is able to synthesize a  policy  within the time limit, and \emph{solved} if the  learned program also matches the user's intent. We determine if a task is {solved} by comparing it against the ground truth program (written manually) and checking if the learned program is semantically equivalent.  Since the manually written policy is intended to work in all environments, tasks  classified as ``solved"  generalize to unseen environments.


\mypar{Running time}
In Figure~\ref{subfig:succes_cactus}, the three  lines indicate the percentage of benchmarks (y-axis) completed with a given time limit (x-axis) for each of the three environments (Easy, Medium, Hard). Across all environments, \tool\  is able to complete 36\% of the tasks within the first 5 seconds, 28\% of the tasks within {6-30} seconds, and 16\% of the tasks within {31-120} seconds. Overall, \tool\ is able to complete 80\% of the tasks within the 2 minute time limit.  

Figure~\ref{fig:results}  also provides a more detailed look at these statistics by showing the percentage of completed tasks with respect to task complexity. In particular, Figure~\ref{subfig:succes_vs_ast} shows the rate of completion  according to the size of the ground truth program.  As expected, the learning problem becomes harder as the complexity of target policy increases. However, even  for the most complex programs, \tool\ is still able to complete the learning task for 68\% of the benchmarks.

\mypar{Generalizability} {As mentioned earlier, completing a task is not the same as ``solving'' it, since the synthesized policy may not match user intent despite being consistent with the demonstrations. We manually inspected all programs synthesized by \tool\  and found that it is equivalent to the ground truth program in {81\%} of the completed cases. Interestingly, {as shown in Figure~\ref{subfig:gen_vs_env}}, we found that \tool's generalization power improves as environment complexity  increases. Intuitively, the more complex the environment, the more objects there are with different properties, so it becomes harder to find multiple programs that ``touch'' exactly the same objects as the demonstration. }

\begin{wrapfigure}{r}{0.45\textwidth}
  \centering
  \includegraphics[width=0.45\textwidth]{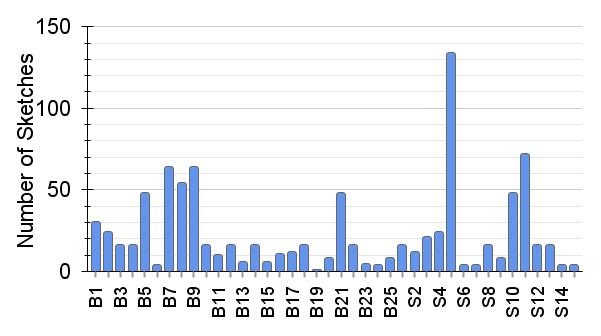}
  \caption{Number of Sketches Considered}
  \label{fig:sketch_stats}
\end{wrapfigure}

\mypar{Search Statistics} 
\new{ 
As explained in Section~\ref{sec:synth}, sketches in \tool\ are enumerated based on increasing order of complexity. In each iteration, a finite number of completions are considered for the current sketch. If none of these completions aligns with the demonstrations, the algorithm advances to the next sketch. This process continues until either a successful match is found or the time limit is reached. Figure~\ref{fig:sketch_stats} presents the average number of sketches encountered before finding the solution or timing out for each task in all three environments. For almost all tasks, \tool\ considers fewer than 50 sketches before termination. One task ($\mathtt{S5}$) is exceptionally challenging and requires 143 sketches before a solution is found.
}

\mypar{Failure analysis} 
As discussed above, there are two reasons why \tool \ may fail to solve a task: (1) it fails to find a policy consistent with the demonstrations within the time limit, or (2) the synthesized policy does not adequately capture the user intent. We have manually inspected both classes of failure cases and report on our findings.

The main cause of \tool's timeouts is due to perception operations. Many of our environments contain a large number of object types, all of which can be arguments of $\stx{scan}$ operations. Our approach tries to overcome this issue by using an LLM to guide search, but in some cases, the LLM proposes the wrong object type to scan for. This causes the synthesizer to go down a rabbit hole, particularly in cases when the proposed object type has many properties associated with it.
We believe more advanced LLMs that can reason about finer-grained properties between the environment and the context of the task can potentially mitigate this issue.

%
We also inspected the cases where \tool \ finds \emph{a} robot execution policy consistent with the demonstrations, but the synthesized policy does not generalize to different environments (i.e., it \textit{completes} the task but fails to \textit{solve} it). There are two main reasons for this, both due to the inadequacy of the demonstrations with respect to the desired task. 
Specifically, if there is only \textit{one} instance of a particular object type in the environment, the synthesizer may not return a program with a $\stx{foreach}$ loop over that object type,  even though the ground truth program contains such a loop. \new{Likewise, 
 \tool\ is capable of inferring conditional blocks only if the demonstrations are \emph{branch complete}. This means that, in the demonstration environment, some instances of a manipulated object type must satisfy the intended property, while others must not. If all (or none) of the object instances of that type are acted upon, the synthesizer cannot learn the conditional block for such manipulations. As noted earlier, this generalizability issue becomes less of a problem in more complex environments with many instances of an object type.
 }

\vspace{2mm}
\noindent\fbox{%
    \parbox{0.99\textwidth}{%
    {\bf RQ1 Summary. }
       Given a 2 minute time limit, \tool\ is able to find a policy consistent with the demonstations for {80\%} of the benchmarks. Furthermore {81\%} of the synthesized programs correspond to the ground truth, meaning that they can generalize to any unseen environment.
    }%
}

\subsection{Ablation Studies} 

As mentioned throughout the paper, there are three key components underlying our  approach, namely (1) learning control flow structures (i.e., sketches), (2) use of LLMs to guide sketch completion, and (3) new technique for proving unrealizability. To better understand the relative importance of each component, we present the results of an ablation study where we disable each component or combinations of components. Specifically, for our ablation study, we consider the following variants of \tool:

\begin{itemize}[leftmargin=*]
    \item {\bf \tool-NoSketch:} This is a variant of \tool\ that does not generate program sketches using regex learning.
    \item {\bf \tool-NoLLM:} This is a variant of \tool\ that does not utilize LLMs for sketch completion.
    \item {\bf \tool-NoPrune:} This is a variant of \tool\ that does not utilize the unrealizability checking procedure (Algorithm~\ref{alg:comp}) for pruning the search space during sketch completion.  
\item {\bf \tool-SketchOnly:} This is a variant of \tool \ that infers control flow sketches (through regex learning) but neither utilizes LLM nor unrealizability checking during sketch completion.
\item {\bf \tool-NoLoopBound:} {This is a variant of \tool\ summarizes loops using the Kleene star operator during the unrealizability checking procedure. That is, it does not reason about the number of loop iterations; however, it is the same as the full \tool \ tool otherwise.}
 \end{itemize}

Figure~\ref{fig:ablation_results} shows the results of our ablation study in the form of a Cumulative Distribution Function (CDF). The x-axis represents the {cumulative} running time, while the y-axis shows the percentage of benchmarks solved in all environments.
The results indicate a significant gap between the number of the tasks solved by \tool\ and its variants defined above. In particular, \tool\ is able to solve 8\% more tasks than \textbf{NoLLM}, 19\% more tasks than \textbf{NoPrune}, 24\% more tasks than \textbf{SketchOnly}, 
and 8\% more tasks than \textbf{NoLoopBound} variants. However, the results for the \textbf{NoSketch} variant are {particularly poor, with \textit{none} of the tasks solved. 

\begin{figure}[t]
    \centering
    \begin{subfigure}[t]{0.49\textwidth}
        \centering
        \includegraphics[height=33mm]{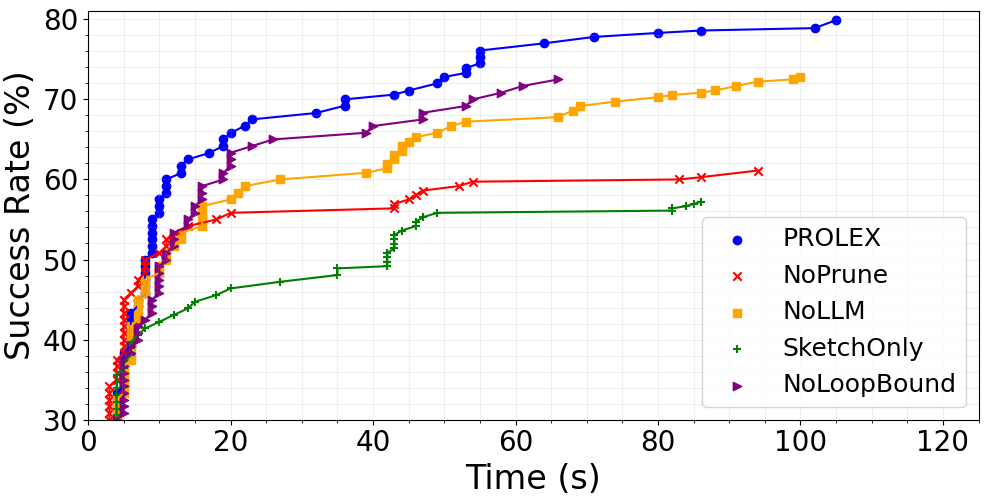}
        \caption{Ablation results over all environments.}
        \label{fig:ablation_results}
    \end{subfigure}
    \begin{subfigure}[t]{0.49\textwidth}
        \centering
        \includegraphics[height=36mm]{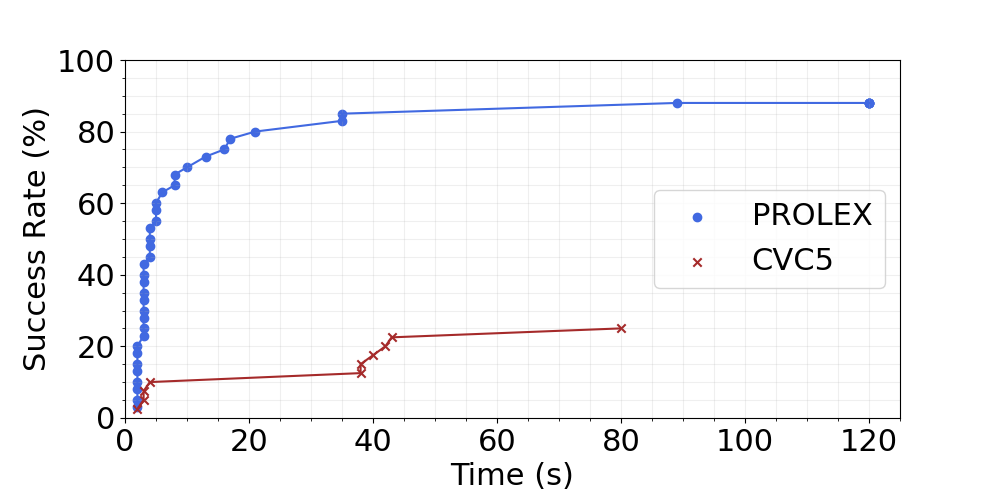}
        \caption{\tool\ vs. CVC5 {in {\bf Easy} environments}.  CVC5 is provided the {\bf ground truth sketch} but \tool \ is not }
        \label{fig:sygus_results}
    \end{subfigure}
    \caption{Experimental Results}
    \label{fig:subfigures}
\end{figure}

\vspace{2mm}
\noindent\fbox{%
    \parbox{\textwidth}{%
    {\bf RQ2 Summary. }
All of the key components of our proposed synthesis algorithm contribute to the practicality of our learning approach. The most important component is regex-based sketch generation, without which none of the tasks can be solved. The unrealizability checking procedure helps solve an additional 18\% of the tasks, and LLM guidance increases success rate by another 8\%. 
    }%
}

\subsection{Comparison with Alternative Approaches}

In this section, we report on our experience comparing \tool \ against alternative approaches. While there is no existing off-the-shelf LfD approach that targets our problem domain (see \secref{bench}), we compare \tool \ against the following two baselines:

\begin{itemize}[leftmargin=*]
    \item {\bf CVC5:} We formulate our learning problem as an instance of syntax-guided synthesis and use a leading SyGuS solver,  CVC5~\cite{DBLP:conf/tacas/BarbosaBBKLMMMN22}, as a programmatic policy synthesizer.
    \item {\bf GPT-Synth:} We use an LLM as a \emph{neural program synthesizer} in our domain. 
    %
    To this end, we consider a baseline called  ``GPT-Synth'' that synthesizes programs in our DSL from demonstrations.
    \end{itemize}


\paragraph{\bf Case Study with CVC5.}

Our programmatic LfD task can be reduced to an instance of the syntax-guided synthesis (SyGuS) problem~\cite{sygus}, which is the standard formulation for synthesis problems. To compare \tool\ against state-of-the-art SyGuS solvers,  we encoded our tasks as instances of SyGuS  and leveraged an off-the-shelf solver, namely,  CVC5~\cite{DBLP:conf/tacas/BarbosaBBKLMMMN22}, which is the winner of the most recent SyGuS competition.

To perform this comparison,
 we defined our DSL
 using the syntactic constraints in SyGuS, and we incorporated semantic constraints  based on the \emph{initial and final environment states} in the demonstrations. 
 {Note that SyGuS solvers are unable to perform synthesis from demonstrations, as demonstrations correspond to intermediate program states, which are not expressible in the SyGuS formulation. Hence, when comparing against CVC5, we only use the initial and final environments and consider a task to be completed if the solver returns a policy that produces the desired environment. Furthermore, we only compare the performance of \tool\ and CVC5 on the ``easy'' environment as encoding the environments in SyGuS requires significant manual labor.}

%


{When we tried to use CVC5 to perform synthesis from scratch (i.e., without a sketch), it was not able to complete any task within a reasonable time limit. Hence, for this evaluation, we manually provided CVC5 with the \emph{ground truth} sketch for the specific task.} The results of this comparison are presented in Figure~\ref{fig:sygus_results}. As in Figure~\ref{fig:ablation_results}, this figure plots  
the cumulative distribution of the percentage of synthesized programs  against  solver time. 
Overall, CVC5 only solves $25\%$ of the tasks, compared to $88\%$ solved by \tool. 
\paragraph{\bf Case Study with GPT-Synth.} For this experiment, we use the GPT 3.5 LLM\footnote{We specifically use the \texttt{text-davinci-003} model, which is the most capable publicly available LLM from OpenAI finetuned for completion, including natural language and code.} to generate programmatic policies from demonstrations.
This study aims to evaluate the effectiveness of LLMs as an end-to-end program synthesizer, henceforth referred to as GPT-Synth.

The prompts used in this study were developed using the "\textit{template}" pattern as described by~\citet{white2023prompt}. This pattern has been shown to be highly effective in situations where the generated output must conform to a specific format that may not be part of the LLM's training data. 
In particular,
for each benchmark task, we created a prompt that includes a description of our DSL, and a set of example demonstrations and the environment states, along with the correct programs for the desired task. 
For a novel task, we provide the language model with the prompt, the demonstrations and environment states for the novel task, and ask it to generate a corresponding program.
%

{To use GPT\,3.5 as an end-to-end synthesizer, we adopt the following methodology, as done in prior work~\cite{chen2021evaluating}. If the first program returned by the LLM  is consistent with the demonstration, GPT-Synth returns that program as the solution. Otherwise, it asks  the model to produce another program, for up to 10 iterations. }
{Unfortunately, we found that GPT-Synth is unable to solve any of the benchmarks  when we provide both the demonstrations and the environment. This behavior seems to be caused  by the large number of entities and relations in the environment -- prior work has reported similar results in other domains (e.g., planning~\cite{mahowald2023dissociating}) where LLMs were used to perform tasks in non-trivial environments.}

\begin{figure}[t]
    \centering
    \begin{subfigure}[t]{0.48\textwidth}
        \centering
        \includegraphics[height=31mm]{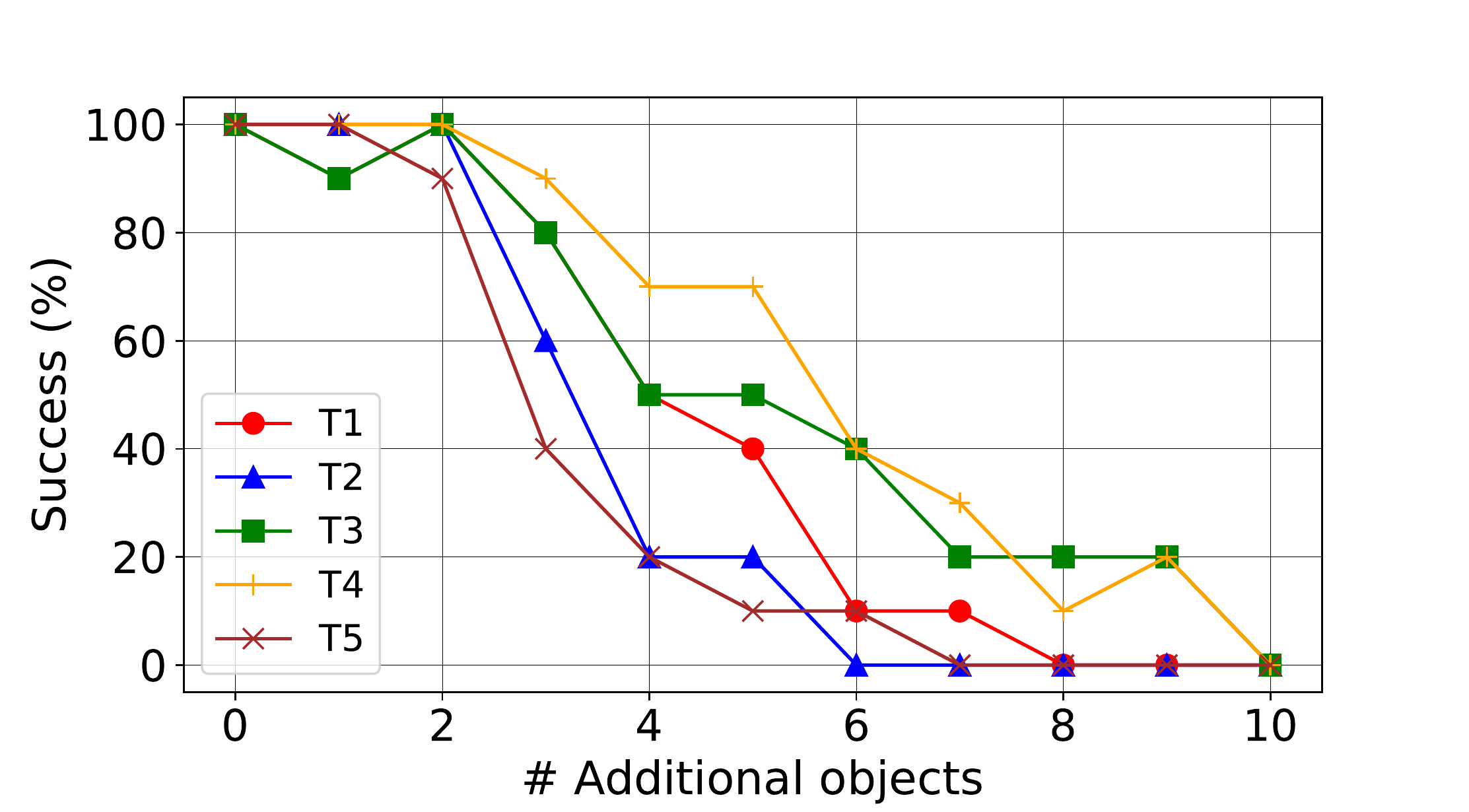}
        \caption{GPT-Synth with environment}
        \label{fig:gpt_synth}
    \end{subfigure}
    \begin{subfigure}[t]{0.48\textwidth}
        \centering
       \includegraphics[height=31mm]{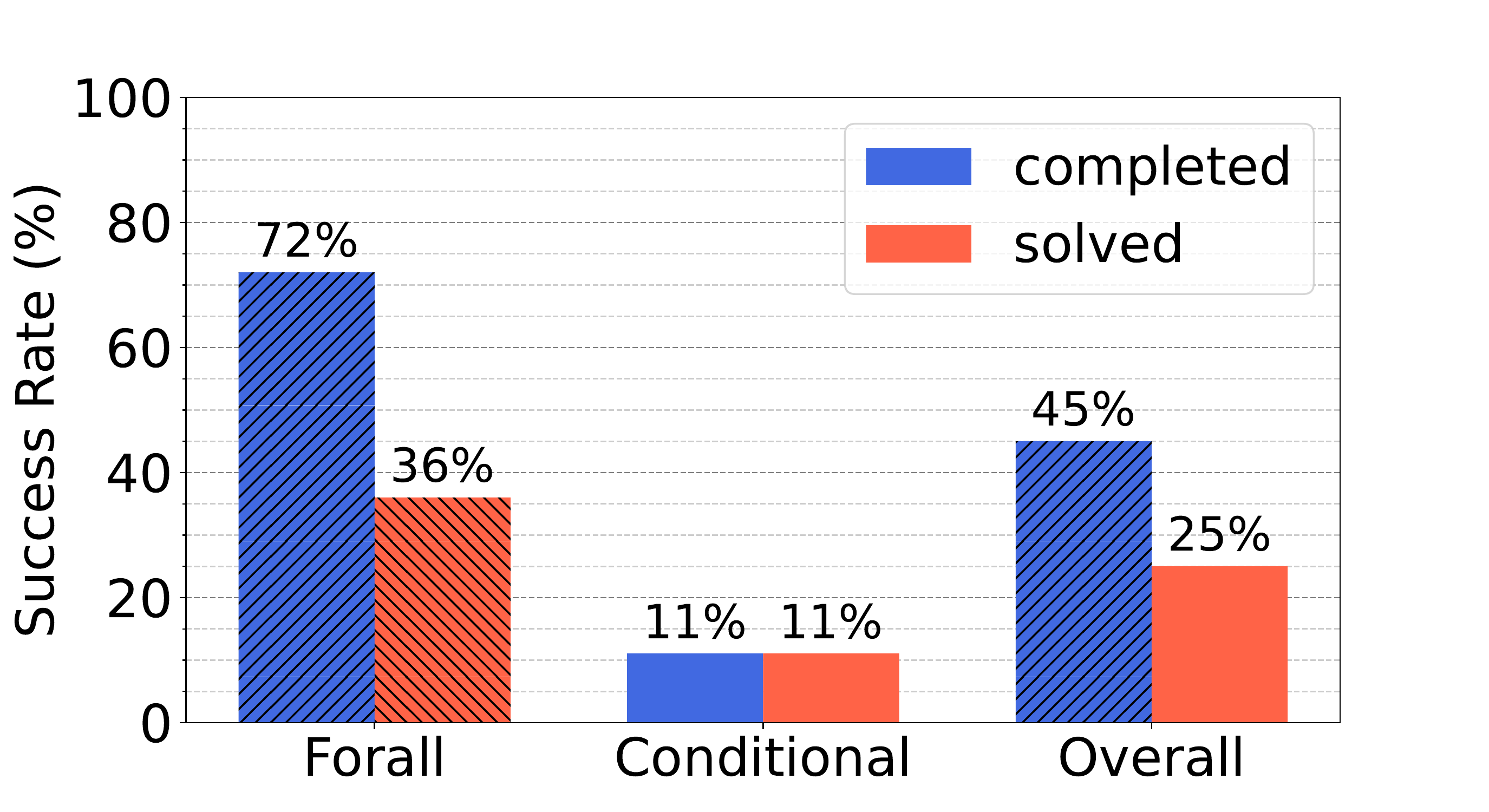} 
        \caption{GPT-Synth without environment}
        \label{fig:gpt_without_env}
    \end{subfigure}
    \vspace{-0.05in}
    \caption{LLM Experiments}
    \label{fig:subfigures}
\end{figure}

{To gain more intuition about how the language model scales with the environment size, we  report on our experience with using GPT-Synth on five representative tasks involving toy environments. We construct these toy environments by incorporating \emph{only} the  objects (and their properties) required for that task plus some additional objects and properties. Figure~\ref{fig:gpt_synth} shows how the success rate of GPT-Synth scales with respect to environment size. Here, the x-axis shows the number of additional objects (and their properties) in the environment and the y-axis shows the success rate. 
As we can see from Figure~\ref{fig:gpt_synth} , GPT-Synth works well if it is given \emph{only} the relevant objects (which is not a realistic usage scenario), but, as environment size increases, its success rate drops dramatically. In fact, when the environment contains only 10 additional objects --- a tiny fraction of our ``Easy" environment ---  the success rate of  GPT-Synth already drops to zero. }

{The reader may wonder if the environment is actually necessary for GPT-Synth to learn the correct program. To answer this question, we  perform an additional experiment where we provide GPT-Synth with only the demonstration, but not the environment. The results of this evaluation are presented in Figure~\ref{fig:gpt_without_env}, where we classify tasks into two categories as ``Forall'' and ``Conditional''. The former class of tasks does not involve branching, whereas the latter does. Here, ``Completed'' shows the percentage of tasks for which GPT-Synth finds a program consistent with the demonstration, and  ``Solved'' shows the percentage of  tasks for which GPT-Synth returns a program that \emph{also} matches the ground truth.} 
{As we can see,  GPT-Synth returns a program consistent with the demonstration for  45\% of all tasks, but it is only able to identify the ground truth program in 25\% of the cases. Furthermore, as one might expect, GPT-Synth is much more effective at the much simpler ``Forall" category of tasks that involve acting on \textit{all} instances of a particular type. In contrast, the ``Conditional" category is much more challenging without having access to the environment, and the success rate of GPT-Synth is only 11\% for this category.  
Intuitively, without knowing which objects have what properties, GPT-Synth has little chance of knowing that there should be a conditional and what its  corresponding predicate should be. We came across a few cases where GPT-Synth is able to ``hallucinate'' the right predicates after several rounds of interaction; but, in general, guessing the user's intent without knowing the environment, is at best a matter of sheer luck.}

\vspace{2mm}
\noindent\fbox{%
    \parbox{\textwidth}{%
    {\bf RQ3 Summary.}
\tool \ performs significantly better than the  CVC5 and GPT-Synth baselines. Even when given the ground truth sketches, CVC5 is only able to return a program consistent with the final environment in 25\% of the cases. On the other hand, the GPT-based synthesizer  cannot solve any tasks when provided with both the demonstration \emph{and} the full environment, but it is  able to solve 25\% of the tasks when it is  given \emph{only} the demonstration. } 
    }%
}

\section{Related Work}
\vspace{-0.07in}
\label{sec:rel}

\mypar{\textbf{Robot Learning from Demonstrations}} 
Our approach builds upon a substantial body of literature on the use of Learning from Demonstration (LfD) techniques to learn robot execution policies~\cite{ARGALL2009469, robotics11060126,4651020}.
%
This literature can be broadly categorized into two approaches: (1) learning neural models to represent robot behaviors~\cite{9117169, doi:10.1177/0278364918770733, 10.5555/1620270.1620297, doi:10.1177/0278364913495721, gail, https://doi.org/10.48550/arxiv.2102.12667,10.5555/1577069.1755839, DBLP:conf/corl/RusuVRHPH17}, and (2) synthesizing programmatic representations of execution policies~\cite{ldips,idips,doi:10.1177/0278364914554471,8794104}.
The most well-established techniques for learning neural models from demonstrations include behavior cloning within the framework of imitation learning~\cite{9117169, gail} and (deep) reinforcement learning (RL) methods~\cite{doi:10.1177/0278364918770733, 10.5555/1620270.1620297, doi:10.1177/0278364913495721}. Empirical studies have demonstrated the efficacy of these neural policies in perception tasks and their ability to perform well in unknown or ill-defined environments. However, such neural models lack robust interpretability and generalization capabilities {--- as a testament to this, there exist no neural LfD algorithms to date capable of leveraging the user demonstrations in the Behavior benchmarks~\cite{pmlr-v164-srivastava22a}}.
The field of transfer learning~\cite{Pan2010ASO,10.5555/1577069.1755839, DBLP:conf/corl/RusuVRHPH17} aims to resolve generalization problems to some degree and also enhance data efficiency. 
{A related setting to our work is applying reinforcement  learning (RL) by specifying the task via formal specification of the goal conditions; in fact, the Behavior benchmark set~\cite{pmlr-v164-srivastava22a} reports the results of such an approach. However, even in the simplest 12 activities, the RL algorithms are unable to complete the tasks even when initiated close to the goal states. Further, even when the actions are abstracted into symbols in Behavior-1K~\cite{pmlr-v205-li23a}, RL approaches demonstrate very poor performance.
These results mainly highlight the complexity of the tasks that we tackle in this paper.

Recently, there has also been  growing interest in developing techniques to enhance the transparency and reliability of RL systems through formal explanations~\cite{10.1609/aaai.v33i01.33012514, interp_RL,interp_RL_krajna,doi:10.1126/scirobotics.aay6276}. These techniques aim to explain different aspects of the learned models, such as inputs and transitions, by finding interpretable representations of neural policies, such as Abstracted Policy Graphs~\cite{10.1609/aaai.v33i01.33012514} or structures in a high-level DSL~\cite{https://doi.org/10.48550/arxiv.1804.02477}.
More recently, there has also been  interest in utilizing  program synthesis methods~\cite{https://doi.org/10.48550/arxiv.2303.01440,Holtz2020RobotAS} to learn robot execution policies from demonstrations as an alternative to neural model learning~\cite{ldips,idips}. These approaches provide improved interpretability, generalizability~\cite{srtr}, and data efficiency. 
%
\tool \ falls into the same class of techniques as these approaches but broadens their applicability in several ways: First, it can learn  policies to handle long-horizon tasks; second, it can synthesize programs with complex control flow, such as loops with nested conditionals and loops; and, third, it can handle environments with a large number of objects and properties.


{
Lastly, recent work proposes a methodology for automated learning of robotic programs for long-horizon human-robot interaction policies from multi-modal inputs, including demonstrations and natural language~\cite{10.1145/3332165.3347957,10.1145/3568162.3576991}. 
The contributions of this work are largely complementary to ours: their main focus is an interface for allowing end-users to draw the navigation path of a robot  on a 2D map of the environment and inferring symbolic traces from this raw path representation. Similar to our regex-learning-based sketch inference algorithm, the synthesizer in this work utilizes an automata learning approach to generalize user-provided traces into Mealy automata, based on the approach in~\cite{Neider2014ApplicationsOA}. However, the control structures in their learned programs  only admit restricted loops with no nesting. Additionally, this synthesizer does not produce loops over objects and locations in the environment and does not address challenges related to perception, i.e. inference of objects not present in the demonstrations. 
}


\mypar{\textbf{Program Synthesis from Demonstrations}} 
This paper is related to a long line of research on program synthesis, which aims to find a program that meets a specified requirement~\cite{synthesis-survey,10.1007/978-3-319-21668-3_10,sketch1, https://doi.org/10.48550/arxiv.1804.01186,sygus,meta-sketch,flashfill,jha2010oracle,FM+18,webqa,relish,vldb20}. 
Different synthesizers adopt different types of specifications, such as input-output examples~\cite{lambda2}, demonstrations~\cite{10.1145/3242587.3242661}, logical constraints~\cite{10.1145/3498682}, refinement types~\cite{synquid}, or a reference implementation~\cite{Wang:2019:Synthesizing}.

Among these, our method is mostly related to the synthesizers that enable programming by demonstration (PbD)~\cite{Lau2003ProgrammingBD,10.1145/3242587.3242661}. Existing PbD 
techniques generalize programs 
either from sequences of user actions~\cite{webrobot,10.1145/3242587.3242661} or sequences of program states~\cite{Lau2003ProgrammingBD}. Our approach is similar to the former, specifically similar to WebRobot~\cite{webrobot}, which can synthesize challenging programs with multiple nested loops from sequences of user actions. However, WebRobot cannot synthesize programs with conditional blocks, which are essential for successfully performing our tasks. Moreover, WebRobot is targeted at web process automation tasks and does not address robotics-related challenges, such as perception and environment complexity, that play a big role in this work.

\mypar{\textbf{Synthesis of Control Structures}}
A main focus of this paper is on the problem of inferring control structures from demonstration traces~\cite{10.1145/872035.872080, 10.1145/2442516.2442529, 1672773}. This is a recognized and challenging problem that is studied in various synthesis methodologies, including approaches for code synthesis from black-box reference implementations~\cite{10.1145/2786805.2786875, 10.1145/1806799.1806833, 10.1145/3586055} and {human-in-the-loop} approaches for program learning~\cite{10.1145/3526113.3545691, Newcomb2019UsingHS, 10.1145/3485530}.
For instance, LooPy~\cite{10.1145/3485530} is a recent human-in-the-loop method that relies on the programmer to act as an oracle and identify properties of consecutive iterations within the body of the target loop. 
%
Unlike these approaches, \tool\ only requires a small number of demonstrations with no additional hints from the end-user.

%
%
%
%
There are relatively few synthesis algorithms that can infer nested loops with  branching, and they typically rely on domain-specific simplifying assumptions to address these challenges.
%
%
%
%
%
%
%
For instance, 
Rousillon~\cite{10.1145/3242587.3242661} is a recent synthesizer that deals with loops and is specifically designed for extracting tabular data from web pages. 
While Rousillon supports nested loops, it can only generate side-effect-free programs intended for information retrieval purposes. 
FrAngel~\cite{10.1145/3290386} is a component-based synthesizer that also handles nested control structures but necessitates users to provide numerous examples, including base and corner cases. 
Lastly, there is a line of work focusing on loop unrolling and rerolling for low-level hardware and software optimization~\cite{10.1145/3591237, 10.1145/3519941.3535072, 10.1109/CGO53902.2022.9741256}, using techniques such as term-rewriting~\cite{10.1145/3385412.3386012}.
We believe that our proposed sketch generation and unrealizability proving technique could be generally useful in any setting that (a) requires synthesizing complex control flow structures and (b) where the algorithm has access to execution traces.

\mypar{\textbf{Reactive Program Synthesis}}
{
This paper is also related to a long line of work on reactive synthesis, where the typical goal is to synthesize finite state machines (FSM)  from temporal specifications~\cite{8206426, 10.1007/978-3-030-59152-6_23,10.1007/978-3-030-53291-8_32,10.1145/3484271.3484972,10.1145/3331545.3342601}.
Traditionally, reactive synthesis has been viewed as a game between two players – the controlled system and its environment – and solving the reactive synthesis problem boils down to finding a winning strategy for the controlled system. The reactive synthesis problem is computationally intractable for general classes of specifications, such as monadic second order logic or full linear temporal logic~\cite{10.1145/75277.75293}, but work by \citet{BLOEM2012911} has shown that this problem can be made tractable by restricting the logic to a subclass known as GR(1) specifications. 
Another successful approach to enhance reactive synthesizers is bounded synthesis, where the number of states of the synthesized implementation is bounded by a constant ~\cite{finkbeiner2018reactive, finkbeiner2013bounded}. 
%
Generally speaking, existing reactive synthesis method differ from our work in three major ways: First, they take as input temporal logic specifications rather than demonstrations, and, second, they are based on deductive rather than inductive synthesis. Third, the focus of reactive synthesis is on synthesizing \emph{reactive systems} represented as finite state machines that take some input from the environment and respond with an output (e.g., action) for a single time step. 
{Thus, while reactive synthesis is well-suited to low-level motor controllers (\eg{} robot social navigation), they are a poor fit for long-horizon tasks, where programs need to reason about actions that may take many time-steps to complete and where the program must relate properties of the initial state to the chosen sequence of long action sequences (\eg{} when a shelf has limited access, deciding to put away the groceries that go at the back of the shelf first, before those in the front).}
%
}

{
Recent work has expanded the scope of reactive synthesis in several ways. 
For example, recent work by \citet{10.1145/3484271.3484972} proposes to combine reactive synthesis with SyGuS (syntax-guided synthesis) to synthesize a broader class of programs that both interact with the environment (i.e., are reactive) and that can also perform data processing. This technique, however, takes as input temporal stream logic (TSL) specifications modulo some background first-order theory. Furthermore, the algorithmic focus of that work is very different in that they show how to combine classical reactive synthesis with SyGuS, whereas our focus is on inductive synthesis for learning generalizable long horizon policies from a small number of demonstrations.
}

{
Another recent related work is \cite{10.1145/3571249}, which introduces a functional reactive synthesis algorithm for learning programs that match a sequence of observed grid frames and corresponding user actions. This work is similar to ours in that it also learns programs from observed traces rather than temporal specifications. They perform synthesis by combining standard functional synthesis techniques with an automata synthesis approach to discover  time-varying latent state in the program. However, the focus of that work is to discover causal mechanisms in Atari-style, time-varying grid worlds. As such, their Autumn DSL is used for specifying how the next state should be computed upon the occurrence of relevant events, whereas our DSL is used to express a sequence of robot actions over a time horizon. As a result, the underlying synthesis techniques are also  different: For example, they use automata learning to discover latent state, whereas we employ automata learning for the entirely different purpose of learning complex control structures.}

{
Finally, reactive functional programs have also been integrated with probabilistic programming features to design simulators for human-robot interaction, enabling the sampling of test scenarios~\cite{9889630}. These simulators generate complex event streams of human actions based on distributions learned from demonstrations. The synthesis strategy employed in this work involves providing a sketch of the target program as input and using probabilistic inference techniques, such as MCMC, to complete the unknown parameters. 
}

 \mypar{\textbf{Enhancing Synthesis using ML Models}} 
Machine learning has proven to be highly effective  for improving   time and accuracy of synthesis~\cite{10.1145/3453483.3454063,10.1145/3571226, 10.1145/3485477,regel2,https://doi.org/10.48550/arxiv.1804.01186,10.1145/3485535}. For example, neural generators trained on partial programs (i.e., sketches) have been shown to accurately  predict the full body of a method from just a few API calls or data types~\cite{https://doi.org/10.48550/arxiv.1703.05698,https://doi.org/10.48550/arxiv.1902.06349,regel}.
In addition, LLMs have been utilized to guide program search~\cite{10.1145/3510003.3510203}. For example, the GPT-3 language model has been applied to mine program components and their distributions for multi-modal program synthesis tasks~\cite{10.1145/3485535}. Our work is similar in approach and leverages an LLM to improve program synthesis. However, to the best of knowledge, \tool\ is the first approach to leverage the LLM's prior knowledge of the semantic relations between real-world entities and actions to guide the search towards reasonable completions.

A related field of research, \textit{neurosymbolic programming}, seeks to combine advances in end-to-end machine learning techniques with program synthesis by leveraging compositional programming abstractions as a means of reusing learned modules across various tasks~\cite{armando_nstransfer,unsuper_learning_eric,Chaudhuri2021NeurosymbolicP,webqa,ns-armando,swarat_PROPEL,mao2018the,pmlr-v119-huang20h,https://doi.org/10.48550/arxiv.2301.03094, 10.1145/3571234}. Because our current approach is based on a symbolic environment representation, it does not require a neurosymbolic DSL. 

 \mypar{\textbf{Program Sketching}}
Program sketches have been introduced as a syntactic framework to guide the generation of candidate programs during a search process. This approach was initially presented in~\cite{sketch1} and has since been widely used~\cite{sketch1,sketch2,sketch3,meta-sketch,webrobot,migrator-pldi19,sqlizer}. While some approaches utilize program sketches that are crafted by the user, others automatically generate a sketch  based on natural language~\cite{regel,sqlizer} or reference implementation~\cite{migrator-pldi19}. Our method also decomposes the synthesis task into two separate sketch generation and sketch completion step but utilizes regex learning to find a sketch that is likely to be a consistent generalization of the user demonstrations.

 \mypar{\textbf{Unrealizability of Program Synthesis}} {
Many prior techniques enhance program synthesis by establishing that a synthesis sub-problem is unrealizable 
~\cite{syngar,neo,10.1145/2993236.2993244,10.1145/1707801.1706338, 10.1007/978-3-319-21401-6_33,10.1007/978-3-030-25540-4_18,10.1145/3236024.3236049}. Existing methods typically rely on domain-specific static analysis and logical reasoning to establish unrealizability  for the task of Programming by Example (PbE). For instance, various approaches have reduced this problem to a SMT instance and leveraged external solvers to find a proof of unrealizability~\cite{10.1007/978-3-030-25540-4_18,10.1145/3434311,neo,morpheus,synquid,10.1145/3519939.3523726}. While the general problem of unrealizability is undecidable~\cite{madhusudan_et_al:LIPIcs:2018:9698}, some recent approaches have used abstract interpretation techniques to establish unrealizability more effectively~\cite{syngar,10.1145/2993236.2993244,regel}. 
Recently, \citet{10.1145/3571216} proposed a Hoare-style reasoning system to formally define, establish, and explain the unrealizability of a problem, aiming to unify existing methods in this domain.
Our synthesis technique also utilizes program abstractions to establish the unrealizability of a search path and prune partial programs. However, to the best of our knowledge, our approach is the first to check compatibility between user demonstrations and partial programs in order to establish unrealizability in the PbD setting.
A different notion of \emph{trace compatibility} has been proposed for synthesis-based transpilation~\cite{10.1145/3527315}; however, that work differs from ours in several ways. First, they define compatibility  between traces of two different programs, while ours is defined between a program and user demonstration. Second, their technique for checking compatibility between traces is very different from ours and relies on a collecting semantics~\cite{CousotC77} of the programming language.
}

\vspace{-0.02in}
\section{Conclusions and Future Work}
\vspace{-0.02in}

{We proposed a new programmatic LfD approach, based on program synthesis,  for learning robot execution policies for long-horizon tasks in complex environments. Our approach first generates a program sketch capturing control flow structure by generating a string abstraction of the given demonstrations and inferring a regular expression that matches those strings. In the second sketch completion phase, our algorithm performs LLM-guided top down search and utilizes a novel procedure for proving unrealizability of partial programs. The latter algorithm for proving unrealizability  can be easily adapted to other PBD settings: the key idea is to generate, via static analysis, a regular expression that  captures all traces of a partial program and then check whether the string representation of the demonstrations belongs to this  language.}

{We have evaluated our implementation, \tool, on 120 benchmarks and show that \tool \ is able to synthesize complex policies with several (nested) loops and conditionals and that it scales to large environments containing thousands of objects and dozens of  distinct objects types. Overall, given a 120 second time limit, \tool \ is able to find a program consistent with the demonstrations for {80\%} of the benchmarks. Furthermore, for {81\%} of the completed  tasks, \tool \ can learn the ground truth program from a single demonstration. To put these numbers in context, we also compare \tool \ against two baselines, including a state-of-the-art SyGuS solver and a neural LLM-based synthesizer, and show that \tool \ significantly outperforms both of them.}

In future work, we are interested in deploying this technique on real robots in physical environments. To this end, we plan to integrate a semantic-aware perception frontend like Kimera~\cite{kimera} to extract the symbolic state of the world as a semantic scene graph, as such a representation would be directly compatible with the \tool{} DSL. We are interested in building a web-based graphical interface to our robots to gather user demonstrations for deployments, building on existing robot deployment management systems like RoboFleet~\cite{robofleet}.

%

\section*{Acknowledgements}
\new{We would like to thank our anonymous reviewers for their helpful and insightful feedback.}
\new{This work was supported in part by NSF Awards \#1762299, \#1918889, \#1901376, \#2046955, and \#2319471, as well as Google, Facebook, Amazon, Intel, and RelationalAI fellowships.}

\bibliography{ref}

\appendix
\section{Supplementary Material}
\label{app:extended_defs}
This section contains the supplementary definitions referenced throughout the paper. 

\subsection{Auxiliary Relation $\rightarrow$}
\label{app:aux_rel}
The auxiliary relation $\rightarrow$ that defines the effect of actions on an environment is given in \autoref{fig:aux}.

\def\horizontalSpace{2mm}
\begin{figure}[h]
\footnotesize
\begin{minipage}[b]{0.66\textwidth}
  \ruleLabel{open}
$$
\RULE{
type({o}) \in \{frdige, drawer, box,\dots\}
\qquad
\intp' = \env.\intp[(openned,{o})\mapsto\top]
}
{
\env \xrightarrow{open,\,{o}}  \env[\intp\mapsto \intp']
}
$$
\end{minipage}
\\[\horizontalSpace]
\begin{minipage}[b]{0.73\textwidth}
  \ruleLabel{put-in}
$$
\RULE{
\env.\intp(open,{o}_2)=\top
\qquad 
\intp' = \env.\intp[(empty,{o}_2)\mapsto\bot][(inside\text{-}of,{o}_1,{o}_2)\mapsto\top]
}
{
\env \xrightarrow{put-in,\, {o}_1,{o}_2} \env[\intp\mapsto\intp']
}
$$
\end{minipage}
\\[\horizontalSpace]
\begin{minipage}[b]{0.72\textwidth}
  \ruleLabel{grab}
$$
\RULE{
o = {type}({o})
\qquad 
\mathtt{objs}' = \env.\objs.remove((\env.\cloc,o),{o}).append(loc_r,{o})

\\
\intp' = \env.\intp[({o},\_,on\text{-}top\text{-}of)\mapsto \bot][({o},\_,inside\text{-}of)\mapsto \bot][\dots]
}
{
\env \xrightarrow{grab,\, {o} }  \env[\intp\mapsto\intp']
}
$$
\end{minipage}

\caption{Auxiliary relation $\rightarrow$ for updating environments following robot actions. These rules are specific to each domain and model the dynamics of the robot's environment. A subset of rules implemented for \tool, modeling a typical household environment, are shown above. Remember that an environment is defined as $\env:=(\locs, \objs, \cloc, \intp)$. A special location $loc_r$ is used for objects that are being carried by the robot.}
\label{fig:aux}
\end{figure}

\subsection{Partial Evaluation}
Rules for partial evaluation of programs on an environment is given in \autoref{fig:partial_eval}.  $\Gamma$ is a partial store mapping a subset of variables to concrete values. $\Gamma,\env\vdash \partial\rightarrow \partial'$ denotes partial evaluation result under $\Gamma$ in environment $\env$. Function $\mathsf{PartialEval}(\partial,\env)$ (referenced in \autoref{alg:feas}) returns $\partial^*$ if and only if $\mathsf{Nil}, \env \vdash \partial \rightarrow \partial^*$. Boolean expression reduction relation $\Downarrow$ used in partial programs is similar to the relation used in \autoref{fig:bool_expr_reduction}, however, the partial variable store ($\Gamma$) replaces the variable complete store ($\sigma$).  
\def\horizontalSpace{5mm}
\begin{figure}[h]
\footnotesize
$$
\RULE{
}{
\Gamma,\env \vdash \rho \rightarrow \rho
}
\qquad
\RULE{
}{
\Gamma,\env \vdash \partial \rightarrow \partial
}
\qquad 
\RULE
{
\Gamma,\env\vdash \partial\rightarrow \partial'' \quad 
\Gamma,\env\vdash \partial'\rightarrow \partial'''
}
{
\Gamma,\env\vdash \partial;\partial'\rightarrow \partial'';\partial'''
}
$$
\\
$$
\RULE{
\tau_o\not\eq ??
\qquad
\cloc\in\mathsf{Domain}(\Gamma) \qquad 
\env.\mathtt{objs}(\Gamma(\cloc),\tau_o) = O 
}{
\Gamma,\env \vdash \stx{scanObj}(\tau_o) \rightarrow O
}
\qquad 
\RULE{
\tau_l\not\eq ?? \qquad \env.\mathtt{locs}(\tau_l) = L
}{
\Gamma,\env \vdash \stx{scanLoc}(\tau_l) \rightarrow L
}
$$
\\
$$
\RULE{
\Gamma,\env \vdash \rho \rightarrow L
\qquad 
v\in\mathsf{Domain}(\Gamma)
\qquad
\Gamma(v) = L[n] 
}{
\Gamma,\env\vdash \stx{let}\ v:= \stx{getNth}(\rho, n) \rightarrow \stx{skip}
}
\qquad 
\RULE{
\phi \Downarrow_{\env,\Gamma} \top
}{
\Gamma,\env \vdash \stx{if}(\phi)\{\partial\} \rightarrow \partial
}
\qquad 
\RULE{
\phi \Downarrow_{\env,\Gamma} \bot
}{
\Gamma,\env \vdash \stx{if}(\phi)\{\partial\} \rightarrow \stx{skip}
}
$$
\\
$$
\RULE{
\Gamma,\env \vdash \rho \rightarrow [i_1,i_2,\dots,i_n]
\qquad 
\forall_{1\leq j \leq n}.\;\partial_j = \partial[v\mapsto i_j]
\qquad
\Gamma,\env \vdash \partial_1;\partial_2;\dots\partial_n \rightarrow \partial'
}{
\Gamma,\env \vdash \stx{foreach}(v\in\rho)\{\partial\}\rightarrow \partial'
}
$$



\caption{Partial Evaluation Rules
}
\label{fig:partial_eval}
\end{figure}

\subsection{Rewrite Rules for the Inferred Regular Expressions}
\label{app:rewrite_rules}
Following is the set of rewrite rules used in \tool's implementation. $x$ and $y$ are meta-variables for any regex:
\begin{itemize}
    \item $(x|y)^*\rightarrow(xy?)^*$
    \item $(x|y)^*\rightarrow(x?y)^*$
    \item $(x|y)^*\rightarrow ((xy)?)^*$
    \item $x^* \rightarrow (x^*)^*$
\end{itemize}

\subsection{Definition of $\mathsf{UpdateAbsEnv}$}
\label{subapp:updateAbsEnv}
In Figure~\ref{fig:updateAbsEnv} we provide domain-specific rules of how atomic action may effect an abstract environment.  We assume $\absenv \xrightarrow[]{a,o} \absenv'$ is defined similarly to the auxiliary relation in Figure~\ref{fig:aux}, i.e., abstract environments too contain an interpretation of relations and properties. Any atomic action will update the abstract environment similar to a concrete environment, however, since variables may refer to a \textit{set} of potential objects or locations, the $\mathsf{UpdateAbsEnv}$ calculates an over-approximation reflecting the effects of action taken on all potential objects. 
In the rules, we assume $\absenv[v]$ represents all possible locations or object instances that were assigned to a variable $v$ as a result of abstract interpretation of a $\stx{scan}$ operation.

\def\horizontalSpace{5mm}
\begin{figure}[h]
\footnotesize
$$
\RULE{
}{
\mathsf{UpdateAbsEnv}(\absenv,\stx{goto}(v)) = \absenv[\mathsf{CurLocs}\mapsto \absenv.\mathsf{CurLocs}\cup\absenv[v]]
}
$$
\\
$$
\RULE{
|\absenv[v]| = n
\qquad
\absenv_0 = \absenv
\qquad
\forall_{0\leq i< n}. \absenv_{i}\xrightarrow[]{a,\absenv[v]_i}\absenv_{i+1}
}{
\mathsf{UpdateAbsEnv}(\absenv,\stx{actUnary}(a,v)) =  \absenv_{n}
}
$$
\caption{Examples of $\mathsf{UpdateAbsEnv}$}
\label{fig:updateAbsEnv}
\end{figure}

\subsection{Proof of Theorem~\ref{th:completeness}}
\label{subapp:proof}
Here we provide the proof of the Theorem~\ref{th:completeness} from Section~\ref{sec:synth}. We begin by restating the theorem. 
Let $\partial$ be an arbitrary partial program and let $\demos$  be the given set of demonstrations. We must prove that for any complete program $P$ that is a completion of $\partial$, the following holds:
\begin{equation}
 P \models \demos \ \Longrightarrow\  \mathsf{Compatible}(\partial, \demos) 
\end{equation}
Based on the definition of the $\mathsf{Compatible}$ function given in \autoref{alg:feas}, we reduce statement (1) above, by factoring out the universal quantifier on demonstrations, to the following:
\begin{equation}
   \forall_{(\env,t)\in\mathcal{D}}.\; (P(\env)=t \ \Longrightarrow\  
\mathsf{Compatible}(\partial, \env, t)) \label{2}
\end{equation}
where the compatibility of $\partial$ to a \emph{single} demo is defined as follows:
\begin{equation}
   \mathsf{Compatible}(\partial, \env, t) := \alpha(t)\in \mathsf{ProgToRegex}(\mathsf{PartialEval(\partial,\env)},\hat{\env}) 
\end{equation}
where $\hat{\env}=\alpha(\env)$, i.e., the abstraction of the environment $\env$ as defined below. As mentioned in \autoref{sec:synth} the abstract environments admit a set of possible current locations for the robot, and drop the interpretation of relations since they are not needed for establishing unrealizability:
$$
\alpha(\env) := (\{\cloc\}, \env.\mathtt{locs}, \env.\mathtt{objs})
$$
Hence, we can reduce the theorem~\ref{th:completeness} to the following:
\begin{equation}
\label{rewritten_theorem}
    \forall_{\env}.\forall_{\partial}.\forall_{P\in\mathsf{Comp}(\partial)}.\;
    \alpha(P(\env)) \in \mathsf{ProgToRegex}(\mathsf{PartialEval(\partial,\env)},\alpha(\env))
\end{equation}
We now construct a proof of (\ref{rewritten_theorem}) by first proving two lemmas. The first lemma states a similar proposition to (\ref{rewritten_theorem}), however, without considering the effects of partial evaluation. The second lemma clarifies the relationship between completions of a partial program $\partial$ and completions of $\mathsf{PartialEval(\partial,\env)}$ for a particular environment $\env$. The two lemmas straightforwardly result in a proof of (\ref{rewritten_theorem}).
\begin{lemma}
\label{l1}
For an arbitrary environment $\env$, partial program $\partial$ and a completion of it, $P\in\mathsf{Comp}(\partial)$, the abstract trace produced by running $P$ on $\env$ is in the language of the over-approximating regular expression of $\partial$. Formally,
$$
\forall_{\env}.\forall_{\partial}. \forall_{P\in\mathsf{Comp}(\partial)}.\; \alpha(P(\env)) \in \mathsf{ProgToRegex(\partial,\alpha(\env))}
$$

\begin{proof}
    We prove this lemma by induction on the derivation tree of $r=\mathsf{ProgToRegex}(\partial,\alpha(\env))$,
 using rules presented in \autoref{fig:partial_to_regex}. 

\paragraph{ATOMIC Rule.}  In this case $\partial$ is assumed to be an atomic action (i.e., $\stx{actUnary}$, $\stx{actBinary}$, or $\stx{goto}$), hence, $\partial$ is already a complete program with no holes and the only completion of $\partial$ is itself, i.e., $\partial=P$. 
Furthermore, based on the rule \textsc{atomic}, the derived regex in this case is $r=\alpha(\delta)$:
\begin{itemize}
    \item If $\partial=\stx{actUnary}(a,v)$ then $\alpha(\partial)=\mathcal{A}_{a,\tau_v}$ and $\alpha(P(\env)) = \mathcal{A}_{a,\tau_v}$
    \item If $\partial=\stx{actBinary}(a,v_1,v_2)$ then $\alpha(\partial)=\mathcal{A}_{a,\tau_{v_1},\tau_{v_2}}$ and $\alpha(P(\env)) = \mathcal{A}_{a,\tau_{v_1}, \tau_{v_2}}$
    \item If $\partial=\stx{goto}(v)$ then $\alpha(\partial)=\mathcal{G}_{\tau_v}$ and $\alpha(P(\env)) = \mathcal{G}_{\tau_v}$
\end{itemize}
In all the above three cases, $\alpha(t)\in r$, hence the sub-proof is complete.

\paragraph{SEQ Rule.} In this case, we assume $\partial=\partial';\partial''$ and complete the sub-proof assuming the validity of the hypothesis on $\partial'$ and $\partial''$. Note that, for any completions $P'$ and $P''$ of $\partial'$ and $\partial''$, their sequential composition, $P';P''$, is a completion of $\partial$. Now assume $r'=\mathsf{ProgToRegex(\partial',\env)} $ and $r''=\mathsf{ProgToRegex(\partial'',\env)}$
Hence, the inductive proof burden is to show that if
\begin{equation}
  P'(\env) = t' \Rightarrow \alpha(t') \in r'
\end{equation}
and 
\begin{equation}
 P''(\env) = t'' \Rightarrow \alpha(t'') \in r''
\end{equation}
then,
\begin{equation}
P';P''(\env) = t \Rightarrow \alpha(t) \in \mathsf{ProgToRegex(\partial';\partial'',\env)} 
\end{equation}
Based on the rule $(\textsc{sequence})$ in \autoref{fig:operational_semantics}, any demonstration trace $t$ accepted by program $P';P''$ can be decomposed to sub-traces $t=t't''$ such that $P'(\env)=t'$ and $P''(\env')=t''$ and $\env'$ is the environment after the execution of $P'$. Hence we can rewrite (7) as follows: 
\begin{equation}
 P';P''(\env) = t';t'' \Rightarrow \alpha(t')\alpha(t'') \in \mathsf{ProgToRegex(\partial';\partial'',\env)} 
\end{equation}
The left hand side of (8) can be rewritten as $\alpha(t')\in r'\wedge \alpha(t'')\in r''$ based on (5) and (6).  
From rule (\textsc{seq}) in \autoref{fig:partial_to_regex} we can infer $\mathsf{ProgToRegex(\partial';\partial'',\env)}=r'r''$. Hence, (8) is rewritten as follows
\begin{equation}
\alpha(t')\in r'\wedge \alpha(t'')\in r'' \Rightarrow \alpha(t')\alpha(t'') \in r'r''
\end{equation}
(9) is true by the definition of regular expressions, hence, the sub-proof is completed.

\paragraph{IF Rule.} Similar to the previous case, note that if $P$ is a completion $\partial$, then $\stx{if}(\_)\{P\}$ is a completion of $\stx{if}(\_)\{\partial\}$. Hence, we assume 
\begin{equation}
P(\env) = t \Rightarrow \alpha(t) \in r 
\end{equation}
where $r=\mathsf{ProgToRegex(\partial,\env)}$, and we show that
\begin{equation}
\stx{if}(\_)\{P\}(\env) = t' \Rightarrow \alpha(t') \in  (r)?
\end{equation}
Depending on the predicate in the conditional, $t'$ is either equivalent to $t$ or $\epsilon$. In either case $t$ is in the language $(r)?$ based on (10) and the definition of regular expressions. The sub-proof is completed.

\paragraph{LET Rule.} This case is trivial: the $\stx{let}$ statement only produces an empty string in the trace which is obviously accepted by the $\epsilon$ regular expression.

\paragraph{LOOP Rule.}
We first assume $\partial$ (the body of the loop $\stx{foreach}(v\in\rho)\{\partial\}$) is over-approximated by $r$, i.e., for any completion $P$ of $\partial$ we have
\begin{equation}
    P(\env)=t \Rightarrow \alpha(t)\in r
\end{equation}
The goal is to show that if the number of loop iterations can be abstracted by a set of natural numbers $\{n_1,n_2,\dots,n_k\}$, then $\stx{foreach}(v\in\rho)\{\partial\}$ is over-approximated by $r^{n_1}|r^{n_2}|\dots|r^{n_k}$. This is also true if for any $i$, $n_i=\star$, in which case, the over-approximating regex is $r^*$, which accepts \emph{any} number of iterations of the loop body. We prove this by a second  induction (on $k$) as follows.

\paragraph{Base Case} In this case, we assume that the body of the loop is executed exactly $n$ times, hence 
$\stx{foreach}(v\in\rho)\{\partial\}$ can be rewritten to $\partial_1;\partial_2;\partial_3;\dots;\partial_n$ where each $\partial_i$ is over-approximate by $r_i$. The rule \textsc{loop} in \autoref{fig:partial_to_regex} necessitates that for each $i$, $r_i=r$, i.e. the regular expression $r$ is an \emph{inductive abstraction} of the loop body. Hence, now we can use the already proved SEQ rule to argue that $\partial_1;\partial_2;\partial_3;\dots;\partial_n$ is over-approximated by $r_1r_2\dots r_n$ which is equivalent to  $r^n$, and the base case of induction is completed.

\paragraph{Inductive Case} We assume the hypothesis holds for $k-1$ and show it also holds for $k$. This can be shown trivially, via an argument similar to the base case and using the definition of the disjunction operator in regular expressions.  
\end{proof}
\end{lemma}

We now introduce and prove another lemma. On a high level, this lemma states that partially evaluating a partial program $\partial$ on environment $\env$ does not eliminate any of the behaviors of completions of $\partial$ on environment $\env$.

\begin{lemma}
\label{l2}
For an arbitrary environment $\env$, partial program $\partial$ and a completion of it, $P\in\mathsf{Comp}(\partial)$, there exists a program $P^*$ that is a completion of $\mathsf{PartialEval(\partial,\env)}$, and $P^*$  produces the same abstract trace as $P$ when executed on $\env$. Formally, 
$$
\forall_{\env}.\forall_{\partial}. \forall_{P\in\mathsf{Comp}(\partial)}.\exists_{P^*\in\mathsf{Comp}(\mathsf{PartialEval}(\partial,\env))}.\;  \alpha(P^*(\env)) = \alpha(P(\env))
$$

\begin{proof}
    By induction on the abstract syntax tree of $\partial$. 

\mypar{Seq Case $(\partial=\partial_1;\partial_2)$}
Induction assumptions are: 
\begin{equation}
\forall_{\env}. \forall_{P_1\in\mathsf{Comp}(\partial_1)}.\exists_{P_1^*\in\mathsf{Comp}(\mathsf{PartialEval}(\partial_1,\env))}.\;  \alpha(P_1^*(\env)) = \alpha(P_1(\env))
\end{equation}
and 
\begin{equation}
\forall_{\env}. \forall_{P_2\in\mathsf{Comp}(\partial_2)}.\exists_{P_2^*\in\mathsf{Comp}(\mathsf{PartialEval}(\partial_2,\env))}.\;  \alpha(P_2^*(\env)) = \alpha(P_2(\env))
\end{equation}
Now we need to derive the following using (13) and (14):
\begin{equation}
\forall_{\env}. \forall_{P\in\mathsf{Comp}(\partial_1;\partial_2)}.\exists_{P^*\in\mathsf{Comp}(\mathsf{PartialEval}(\partial_1;\partial_2,\env))}.\;  \alpha(P^*(\env)) = \alpha(P(\env)).
\end{equation}
We first Skolemize the existential quantifier in (13) and (14) and introduce a function $\mathcal{F}_1^{P,\env}$ which returns the corresponding program $P_1^*\in\mathsf{Comp}(\mathsf{PartialEval(\partial_1,\env)})$, and similarly, we introduce $\mathcal{F}_2^{P,\env}$ function for $P_2^*\in\mathsf{Comp}(\mathsf{PartialEval(\partial_2,\env)})$. Now we can compose $\mathcal{F}_1^{P,\env}$ and $\mathcal{F}_2^{P,\env}$ sequentially and apply it as a Skolem function and prove (15).

\mypar{Let Case $(\partial=\stx{let}\; v:= \stx{getNth}(\rho, n))$}
Based on the syntax of partial programs, we have
$$\rho\in\{\stx{scanObj}(\tau_o), \stx{scanLoc}(\tau_l), \stx{scanObj}(??), \stx{scanLoc}(??)\}$$
For each of the above four possibilities for $\rho$, any completion of $\partial$, or similarly, any completion of $\mathsf{PartialEval}(\partial,\env)$ only produce the empty string as their abstract trace. Hence, the induction case is proved trivially. 

\mypar{Atomic Cases} Based on the syntax of partial progtams, we have 
$$\partial \in \{\stx{actUnary}(a,\_), \stx{actBinary}(a,\_,\_), \stx{goto}(\_), \stx{skip} \}$$
where $\_$ can be either a variable ($v_i$) or a typed hole ($??:\tau$). In all possible cases, based on the rules in \autoref{fig:partial_eval} $\partial$ does not change, i.e., $\mathsf{PartialEval}(\partial,\env)=\partial$ and induction case is trivially proved.

\mypar{If Cases ($\partial\in\{\stx{if}(\phi)\{\partial'\}, \stx{if}(??)\{\partial'\}  \}$)}
We prove this sub-case by noting that the Boolean reduction relation for partial evaluation $(\Downarrow_{\env,\Gamma})$ is consistent with the Boolean reduction relation  used in the operational semantics of \tool\ programs $(\Downarrow_{\env,\sigma})$, given in \autoref{fig:bool_expr_reduction}. Also note that partial evaluation does not affect $\partial$ if the predicate is unknown. As a result, we consider two scenarios where $\phi\Downarrow_{\env,\Gamma}\top$ and $\phi\Downarrow_{\env,\Gamma}\bot$. In the former case, the partial evaluation rules rewrite $\partial$ with $\partial'$ and in the former case with $\stx{skip}$. This is consistent with rules \textsc{if-t} and \textsc{if-f}, hence it is straightforward to show that traces generated by completions of $\partial$ and $\partial'$ using an inductive argument on the length of the demonstration trace. In the later case, the only trace generated by any completion of $\partial$ and $skip$ is $\epsilon$ and the inductive case is proved.

\mypar{Loop Cases ($\partial=\stx{foreach}(v\in\rho)\{\partial'\}$)} Similar to the previous cases, we consider two scenarios where $\partial$ is affected by partial evaluation and when it is not. In the former case is the lemma is trivially true, using a similar argument as above. In the later case, $\rho$ is evaluated to $L$, a list of concrete object or location instances during partial evaluation. The last rule in \autoref{fig:partial_eval} states that in this case the loop is unrolled and  $\partial$ is rewritten as sequential composition of each iteration per each object or location. Hence, $\partial=\partial_1,\partial_2,\partial_3,\dots,\partial_n$. We can now apply the already proved sequential case above to derive the validity of the lemma for $\partial$ based on validity of it for each of $\partial_i$.
    
\end{proof}
\end{lemma}

We now complete the proof of  (\ref{rewritten_theorem}) using Lemma~\ref{l1} and Lemma~\ref{l2}. Let's first Skolemize the existential quantifier in Lemma~\ref{l2} and introduce a function $\mathcal{F}^{P,\env}$ that returns $P^*$, i.e. the corresponding program to $P$ in the space of completion of $\mathsf{PartialEval}(\partial,\env)$. Hence, Lemma~\ref{l2} can be rewritten as follows: 
\begin{equation}
    \label{A}
    \forall_{\env}.\forall_{\partial}.  \forall_{P\in\mathsf{Comp}(\partial)}.\; \alpha(\mathcal{F}^{P,\env}(\env)) = \alpha(P(\env))
\end{equation}
We now rewrite (\ref{rewritten_theorem}) using the above universal equality:
\begin{equation}
\label{rewritten_theorem2}
    \forall_{\env}.\forall_{\partial}.\forall_{P\in\mathsf{Comp}(\partial)}.\;
     \alpha(\mathcal{F}^{P,\env}(\env)) \in \mathsf{ProgToRegex}(\mathsf{PartialEval(\partial,\env)},\alpha(\env))
\end{equation}
Now, since we know that $\mathcal{F}^{P,\env}(\env)\in \mathsf{Comp}(\mathsf{PartialEval}(\partial,\env))$, 
we can derive  (\ref{rewritten_theorem2}) directly from Lemma~\ref{l1}, by plugging  $\mathcal{F}^{P,\env}$ as $P$ and $\mathsf{PartialEval}(\partial,\env)$ as $\partial$. Hence, the proof of \autoref{th:completeness} is completed.

%

%

\subsection{Description of Benchmark Tasks}
\label{app:bench_task_desc}
A short description of each benchmark task is presented in \autoref{table:task_nl}. 
\begin{table}[h]
\centering
  \caption{\textsc{prolex-ds} benchmark set of long-horizon robotic tasks}
\begin{footnotesize}
  \begin{tabular}{|p{0.03\linewidth}|p{0.78\linewidth}|l|}  
      \hline
     Task & Short Description & Source\\
 \hline
    B1 & Sort books by colour into corresponding drawers & Behavior\\
    B2 & Grab fruit from the fridge and place them on the counter & Behavior\\
    B3 & Box all books up for storage & Behavior\\
    B4 & Bring wood into the kitchen and place it near the fridge & Behavior\\
    B5 & Brush lint off of all the clothing and place them in the drawer & Behavior\\
    B6 & Close all the open doors in the house &Behavior\\
    B7 & Grab all bottles and matchboxes and put them in the living room garbage &  Behavior \\
    B8 & Clean the plates and cups on the table and push in all the chairs & Behavior \\
    B9 & Store the recently bought groceries in their proper locations &  Behavior\\
    B10 & Grab the lamps near the door and place them near the bed & Behavior \\
B11	& go to the kitchen, grab the brush, clean the stove&	Behavior\\
B12&	go to the kitchen put the dirty Plates in the dishwasher &Behavior\\
B13&	go to the kitchen and clean all Plates		&Behavior\\
B14&	go to the kitchen, put all empty mugs into the sink	&Behavior\\
B15&	go to the kitchen, clean all fruits	&Behavior\\
B16&	go to the kitchen, grab a cleaning tool, clean all windows	&Behavior\\
B17	&go to the bedroom, put all clothes into the drawer	&Behavior\\
B18&	go to the kitchen, grab all plates and put them inside the sink		&Behavior\\
B19	&go to the livingroom, pickup all trash, take it to the kitchen		&Behavior\\
B20&	Go to the kitchen, clean the floor		&Behavior\\
B21&	go to the kitchen, open the fridge grab spoiled fruits, put them in the sink	&Behavior\\
B22&	go to the bedroom, grab all lights that are off, put them on bed	&Behavior\\
B23&	go to all rooms, close all doors		&Behavior\\
B24&	go to all rooms, close all windows		&Behavior\\
B25&	go to each room, grab all bottles, go to the kitchen, empty all bottles		&Behavior\\
    S1 & Grab all the dry clothes from the laundry room and put them in a basket  & Survey \\
    S2 & Grab the towel in the kitchen and clean all the chairs in all the rooms & Survey \\
    S3 & Grab all the clothes in the living room, put them in the basket and bring the basket to the bedroom  & Survey \\
    S4 & Grab all the matchboxes in the living room and put them in the drawer  & Survey \\
    S5 & Grab all plates and mugs from the dishwasher. Put the plates in the sink, put the mugs in the drawer & Survey \\
    S6 & Turn off all lamps in all rooms  & Survey \\
    S7 & Pickup the basket in the living room and grab all the pillows on all beds & Survey \\
    S8 & Grab all red books and put them on the living room table & Survey \\
    S9 & Grab all the boxes near the living room door and put them on the table & Survey \\
    S10 & Grab the basket in the living room and put it next to the bed in the bedroom. Then grab all the clothes in the bedroom and put them in the basket  & Survey \\
    S11 & Grab all vegetables and fruits from inside the fridge and place them in the garbage bin & Survey \\
    S12 & Open all windows in all rooms, then sweep all the floors, then close all the windows  & Survey\\
    S13 & Grab an empty box from the living room, go to all the bedroom and grab all the books on the bed and put them in the box. Then take the box back to the living room  & Survey\\
    S14 & Open the kitchen window if the stove is on  & Survey \\
    S15 & Go to each bedroom and grab a pillow from the cupboard and put it on the bed & Survey \\
 \hline
\end{tabular}
\end{footnotesize}
  \label{table:task_nl}
\end{table}

\subsection{Main Synthesis Results}
The details of the ground truth programs and the synthesis results per each task and environment is presented in \autoref{table:detailed_results}.
\begin{table}[h]
\centering
  \caption{Details of the Ground Truth Programs and the Main Synthesis Results}
  \label{table:detailed_results}
\begin{scriptsize}
  \begin{tabular}{|p{0.03\linewidth}|p{2mm}|p{2mm}|p{2mm}|l|l|l|l|l|l|l|l|l|l|}  
      \hline
    \multirow{2}{*}{ ID } & \multirow{2}{*}{\#L}& \multirow{2}{*}{\#C }& \multirow{2}{*}{\#V} &\multirow{2}{*}{Size} &  \multicolumn{3}{c|}{{Easy Environment}}
    &  \multicolumn{3}{c|}{{Medium Environment}} &  \multicolumn{3}{c|}{{Hard Environment}}
    \\
    \cline{6-14}
    & & & & & Comp & Solve & Time (s)& Comp & Solve & Time (s) & Comp & Solve & Time (s)
\\
 \hline
    B1 &  1 & 2 & 3 & 50 & Yes & Yes & 11 & Yes & Yes & 36 & No & No & T/O \\ 
    B2 & 1 & 0 & 3 & 36 & Yes & Yes & 4 & Yes & Yes & 6 & Yes & Yes & 22 \\ 
    B3 &  1 & 0 & 2 & 23 & Yes & Yes & 3 & Yes & Yes & 4 & Yes & Yes & 11 \\ 
    B4 &  2 & 0 & 2 & 32 & Yes & Yes & 4 & Yes & Yes & 8 & Yes & Yes & 55 \\ 
    B5 &  1 & 0 & 4 & 54 & Yes & Yes & 8 & Yes & Yes & 10 & Yes & Yes & 32 \\ 
    B6 &  2 & 1 & 1 & 16 & Yes & Yes & 2 & Yes & Yes & 5 & Yes & Yes & 19 \\ 
    B7 &  3 & 0 & 3 & 52 & Yes & Yes & 9 & Yes & Yes & 36 & No & No & T/O \\ 
    B8 &  3 & 3 & 3 & 51 & No & No & T/O & No & No & T/O & No & No & T/O \\ 
    B9 &  2 & 0 & 4 & 56 & Yes & No & 6 & Yes & Yes & 13 & Yes & Yes & 80  \\ 
    B10 & 1 & 1 & 3 & 35 & Yes & No & 8 & Yes & No & 53 & Yes & No & 105 \\ 
    B11	&	1&	0&	3 & 21 & Yes & Yes & 2 & Yes & Yes & 3 & Yes & Yes & 6   \\
    B12&	1	&1	&3 & 28 & Yes & No & 3 & Yes & No & 4 & No & No & T/O	\\ 
    B13&	1	&0	&2 & 11 & Yes & Yes & 2 & Yes & Yes & 2 & Yes & Yes & 5	\\ 
    B14&	1	&1	&3 & 28 & Yes& No & 3& Yes & No & 43 & No & No & T/O	\\ 
    B15&	1	&0	&2 & 11 & Yes& Yes &2 & Yes & Yes & 2 & Yes & Yes & 5	\\ 
    B16&	1	&0	&3 & 21 & Yes& No& 2& Yes & Yes & 3 & Yes & Yes & 6	\\ 
    B17	&1	&0	&3 &29 & Yes & Yes & 3 & Yes & Yes & 4 & Yes & Yes & 9	\\ 
    B18&1	&0	&3 &23 & Yes&Yes &3 & Yes & Yes & 5 & Yes & Yes & 19	\\ 
    B19		&1	&0	&2 &13 & No&No &T/O & No & No & T/O & No & No & T/O	\\ 
    B20&	0	&0	&3 &22 & Yes&Yes &2 & Yes & Yes & 3 & Yes & Yes & 4	\\ 
    B21&1	&1	&4 &38 & Yes & Yes & 9 & Yes & Yes & 13 & Yes & Yes & 50	\\ 
    B22&	1	&1	&3 &28 & Yes & Yes & 8 & Yes & Yes & 53 & Yes & No & 102	\\ 
    B23&	1	&0	&2 &12 & Yes & Yes & 2 & Yes & Yes & 3 & Yes & Yes & 9	\\ 
    B24&	1	&0	&2 &12 & Yes & No & 2 & Yes & Yes & 3 & Yes & Yes & 9	\\ 
    B25&	2	&0	&3 &23 & Yes & Yes & 3 & Yes & Yes & 8 & Yes & Yes & 71	\\ 
    S1 &  1 & 1 & 2  &27 & Yes & Yes & 6 &Yes & Yes & 10 & Yes & Yes & 55 \\ 
    S2 &  2 & 0 & 2  &21 & Yes & Yes & 3 & Yes & Yes & 4 & Yes & Yes & 14 \\ 
    S3 &  1 & 0 & 2 &46 & Yes & No & 4 & Yes & Yes & 6 & Yes & Yes & 23 \\ 
    S4 &  1 & 0 & 2  &26 & Yes & Yes & 3 & Yes & Yes & 4 & Yes & Yes & 11 \\ 
    S5 &  2 & 2 & 4  & 72& No & No & T/O & No & No & T/O & No & No & T/O \\ 
    S6 &  2 & 0 & 1  &12 & Yes & Yes & 2 & Yes & Yes & 3 & Yes & Yes & 10 \\ 
    S7 &  2 & 1 & 3  &40 & No & No & T/O & No & No & T/O & No & No & T/O \\ 
    S8 &  2 & 1 & 2  &33 & Yes & Yes & 7 & Yes & Yes & 55 & No & No & T/O \\ 
    S9 &  1 & 1 & 3  &35 & Yes & No & 2 & Yes & No & 3 & Yes & No & 5\\ 
    S10 &  1 & 0 & 3  &40 & Yes & Yes & 7 & Yes & Yes & 9 & Yes & Yes & 17 \\ 
    S11 &  2 & 1 & 4  &45 & Yes & Yes & 49 & Yes & Yes & 64 & No & No & T/O \\ 
    S12 &  6 & 0 & 3  &38 & Yes & Yes & 4 & Yes & Yes & 8 & Yes & Yes & 20 \\ 
    S13 &  2 & 2 & 3  &61 & Yes & No & 5 & Yes & No & 45 & Yes & No & 86\\ 
    S14 &  0 & 1 & 2  &23 & Yes & No & 2 & Yes & No & 2 & Yes & No & 3 \\ 
    S15 &  1 & 0 & 3  &35 & No & No & T/O & No & No & T/O & No & No & T/O \\ 
 \hline
\end{tabular}
\end{scriptsize}
\end{table}

\end{document}